\documentclass[a4paper, 11pt,reqno]{amsart}
\usepackage{bbding}
\usepackage{tipa}
\usepackage{pifont}
\usepackage{amsfonts}
\usepackage{mathrsfs}

\usepackage[english]{babel}
\usepackage{pgf,pgfarrows,pgfnodes,pgfautomata,pgfheaps}
\usepackage{amsmath,amssymb}
\usepackage[latin1]{inputenc}
\usepackage{dsfont}
\usepackage{multicol}

\newtheorem{theorem}{Theorem}[section]
\newtheorem{definition}[theorem]{Definition}
\newtheorem{lemma}[theorem]{Lemma}

\newtheorem{corollary}[theorem]{Corollary}

\newtheorem{exmp}[theorem]{Example}

\begin{document}
\title{Bounds and Constructions for $\overline{3}$-Separable Codes with Length $3$}
\author{Minquan Cheng}
\address{Minquan Cheng: Information Security and National Computing Grid Laboratory, Southwest Jiaotong University, Chengdu 610031, China}
\email{chengqinshi@hotmail.com}
\author{Jing Jiang}
\address{Jing Jiang: School of Computer Science and Information Technology, Guangxi Normal University, Guilin 541004, China}
\email{jjiang2008@hotmail.com}
\author{Haiyan Li}
\address{Haiyan Li: School of Mathematics and Statistics, Guangxi Normal University, Guilin 541004, China}
\email{lhyqw2015@sina.com }
\author{Ying Miao}
\address{Ying Miao: Faculty of Engineering, Information and Systems, University of Tsukuba, Tsukuba, Ibaraki 305-8573, Japan}
\email{miao@sk.tsukuba.ac.jp}
\author{Xiaohu Tang}
\address{Xiaohu Tang: Information Security and National Computing Grid Laboratory, Southwest Jiaotong University, Chengdu 610031, China}
\email{xhutang@home.swjtu.edu.cn}

\thanks{Cheng is supported in part by NSFC (No.11301098),
Guangxi Natural Science Foundations (No.2013GXNSFCA019001 and 2014GXNSFDA118001),
and Foundation of Guangxi Education Departmen1t (No.2013YB039).
Jiang is supported by the Guangxi Natural Science Foundation (No.2012GXNSFGA060004)
and Guangxi ``Bagui Scholar$"$ Teams for Innovation and Research.
Miao is supported by JSPS Grant-in-Aid for Scientific Research (C) (No.15K04974).}



\date{Received: date / Accepted: date}

\maketitle

\begin{abstract}
Separable codes were introduced to provide protection against illegal redistribution of copyrighted multimedia material.
Let $\mathcal{C}$ be a code of length $n$ over an alphabet of $q$ letters.
The descendant code ${\sf desc}(\mathcal{C}_0)$ of $\mathcal{C}_0 = \{{\bf c}_1, {\bf c}_2, \ldots, {\bf c}_t\} \subseteq {\mathcal{C}}$
is defined to be the set of words ${\bf x} = (x_1, x_2, \ldots,x_n)^T$ such that $x_i \in \{c_{1,i}, c_{2,i}, \ldots, c_{t,i}\}$ for all $i=1, \ldots, n$,
where ${\bf c}_j=(c_{j,1},c_{j,2},\ldots,c_{j,n})^T$.
$\mathcal{C}$ is a $\overline{t}$-separable code if for any two distinct $\mathcal{C}_1, \mathcal{C}_2 \subseteq \mathcal{C}$
with $|\mathcal{C}_1| \le t$, $|\mathcal{C}_2| \le t$, we always have ${\sf desc}(\mathcal{C}_1) \neq {\sf desc}(\mathcal{C}_2)$.
Let $M(\overline{t},n,q)$ denote the maximal possible size of such a separable code.
In this paper, an upper bound on $M(\overline{3},3,q)$ is derived by considering an optimization problem related to a partial Latin square,
and then two constructions for $\overline{3}$-SC$(3,M,q)$s are provided by means of perfect hash families and Steiner triple systems.
\end{abstract}

\section{Introduction}    %
\label{pre}                     %

Separable codes can be used to construct multimedia fingerprinting codes which can effectively trace and even identify
the sources of pirate copies of copyrighted multimedia data, see, {\it e.g.}, \cite{C.M,T.W}.
They are of interest in combinatorics and can be also used to study the classic digital fingerprinting codes such as
identifiable parent property (IPP) codes \cite{HLLT, S.S.W}, frameproof codes (FPCs) \cite{Bl,B.D}, perfect hash families (PHFs) \cite{STW} and so on.
Cheng and Miao \cite{C.M} pointed out that IPP codes, FPCs, PHFs and some other structures in digital fingerprinting are
in fact examples of separable codes with additional properties.

Let $n,M$ and $q$ be positive integers, and $Q$ be an alphabet with $|Q|=q$.
A set $\mathcal{C} = \{{\bf c}_1,{\bf c}_2,\ldots, {\bf c}_M\} \subseteq Q^n$ is called an $(n,M,q)$ code and each ${\bf c}_i$ is called a codeword.
Without loss of generality, we may assume $Q=\{0,1,\ldots,q-1\}$. When $Q=\{0,1\}$, we also use the word ``binary".
Given an $(n,M,q)$ code, its incidence matrix is the $n\times M$ matrix on $Q$ in which the columns are the $M$ codewords in $\mathcal{C}$.
We do not distinguish an $(n,M,q)$ code and its incidence matrix unless otherwise stated.

For any subset of codewords $\mathcal{C}_0\subseteq \mathcal{C}$,
we define the set of $i$th coordinates of $\mathcal{C}_0$ as
$$\mathcal{C}_0(i)=\{c_i \in Q \ | \ {\bf c} = (c_1,c_2, \ldots, c_n)^T  \in \mathcal{C}_0\}, \  \ 1 \le i \le n, $$
and the descendant code of $\mathcal{C}_0$ as
\[ {\sf desc}(\mathcal{C}_0)=\{{\bf x}=(x_1,x_2,\ldots,x_n)^T  \in Q^n \ | \ x_i \in \mathcal{C}_0(i), \ 1 \le i \le n\},\]
that is,
\[{\sf desc}(\mathcal{C}_0)=\mathcal{C}_0(1) \times \mathcal{C}_0(2) \times \cdots \times \mathcal{C}_0(n).\]

\begin{definition} \rm
\label{d111}
Let $\mathcal{C}$ be an $(n,M,q)$ code and $t \ge 2$ be an integer.
\begin{itemize}
\item[(1)] $\mathcal{C}$ is a $t$-separable code, or $t$-SC$(n,M,q)$,
if for any $\mathcal{C}_1, \mathcal{C}_2 \subseteq \mathcal{C}$ such that
$|\mathcal{C}_1| = |\mathcal{C}_2| = t$ and $\mathcal{C}_1 \neq \mathcal{C}_2$,
we have ${\sf desc}(\mathcal{C}_1) \neq {\sf desc}(\mathcal{C}_2)$,
that is, there is at least one coordinate $i$, $1 \le i \le n$, such that $\mathcal{C}_1(i) \neq \mathcal{C}_2(i)$.
\item[(2)] $\mathcal{C}$ is a $\overline{t}$-separable code, or $\overline{t}$-SC$(n,M,q)$,
if for any $\mathcal{C}_1, \mathcal{C}_2 \subseteq \mathcal{C}$ such that
$|\mathcal{C}_1| \le t$, $|\mathcal{C}_2| \le t$ and $\mathcal{C}_1 \neq \mathcal{C}_2$,
we have ${\sf desc}(\mathcal{C}_1) \neq {\sf desc}(\mathcal{C}_2)$,
that is, there is at least one coordinate $i$, $1 \le i \le n$, such that $\mathcal{C}_1(i) \neq \mathcal{C}_2(i)$.
\item[(3)] $\mathcal{C}$ is a $t$-frameproof code, or $t$-FPC$(n,M,q)$,
if for any $\mathcal{C}' \subseteq \mathcal{C}$ such that $|\mathcal{C}'| \le t$,
we have that ${\sf desc}(\mathcal{C}') \bigcap \mathcal{C} = \mathcal{C}'$, that is,
for any ${\bf c} =(c_1,\ldots, c_n)^{T} \in \mathcal{C} \setminus \mathcal{C}'$,
there is at least one coordinate $i$, $1 \le i \le n$, such that $c_i \not\in \mathcal{C}'(i)$.
\end{itemize}
\end{definition}

Cheng et al. \cite{C.J.M,C.M} established the following relationships between frameproof codes and separable codes.

\begin{lemma} \rm (\cite{C.M})
\label{le28-5}
A $t$-FPC$(n,M,q)$ is also a $\overline{t}$-SC$(n,M,q)$.
\end{lemma}

\begin{theorem} \rm (\cite{C.J.M})
\label{adle1}
An $(n,M,q)$ code $\mathcal{C}$ is a $\overline{t}$-SC$(n,M,q)$ if and only if
$\mathcal{C}$ is a $(t-1)$-FPC$(n,M,q)$ and a $t$-SC$(n,M,q)$.
\end{theorem}

Since the parameter $M$ of a $\overline{t}$-SC$(n,M,q)$ corresponds to the number of fingerprints
assigned to authorized users who purchased the right to access the copyrighted multimedia data,
we should try to construct separable codes with $M$ as large as possible, given length $n$.
Let $M(\overline{t},n,q) = \hbox{max} \{M \ | \ \hbox{there exists a } \overline{t} \hbox{-SC}(n,M,q)\}$.
A $\overline{t}$-SC$(n,M,q)$ is said to be optimal if $M = M(\overline{t},n,q)$.
Similarly, a $t$-FPC$(n,M,q)$ is optimal if $M$ is the largest possible value given $n$, $q$ and $t$.
From the relationship above, Cheng et al. obtained the following upper bound on $M(\overline{t},n,q)$.

 \begin{theorem} \rm(\cite{C.J.M})
 \label{thupper1}
Let $n,q$ and $t$ be positive integers such that $t \ge 3$ and $n \ge 2$,
and let $r \in \{0,1,\ldots,t-2\}$ be the remainder of $n$ on division by $t-1$.
If $M(\overline{t},n,q) > q$, then
$$ M(\overline{t},n,q) \le {\max}\{q^{\lceil n/(t-1) \rceil}, r(q^{\lceil n/(t-1) \rceil}-1)+(t-1-r)(q^{\lfloor n/(t-1) \rfloor}-1)\}.$$
\end{theorem}

Bazrafshan and Trung showed the following results.

\begin{lemma} \rm (\cite{B.M})
\label{le29-5}
For any positive integers $t$ and $q\geq 2$, the following code 
is an optimal $t$-FPC$(n,n(q-1),q)$ when $2\leq n\leq t$.
 \begin{eqnarray*}
 \label{eq00}
\left(\begin{array}{cccccccccccccc}
1\ \ & 2\ \   & \ldots  &q-1&\  0\ \  & 0\ \  & \ldots& 0\ \ &\ldots &0\ \ & 0\ \ &\ldots&0 \\
0\ \ & 0\ \  & \ldots  &0 \ \ &\ 1\ \ & 2\ \ & \ldots  &q-1& \ldots&0\ \ &0\ \ &\ldots&0 \\
&&\vdots&&&&\vdots&&\ddots&&&\vdots&\\
0 \ \ & 0\ \  & \ldots  &0\ \  &\ 0\ \ & 0\ \  & \ldots  &0\ \ &\ldots & 1\ \ & 2\ \ & \ldots  &q-1
\end{array}\right)
\end{eqnarray*}
\end{lemma}

\begin{lemma} \rm (\cite{B.M})
\label{le4-8}
For a $t$-FPC$(n,M,q)$ with $n=1+t$ we have\\[0.2cm]
\indent (i)  $M \leq q^2$, if $n\leq q$,\\[0.2cm]
\indent (ii)  $M \leq nq$, if $n > q$.
\end{lemma}

According to Theorem \ref{adle1} and Lemmas \ref{le28-5}, \ref{le29-5}, \ref{le4-8}, the following corollaries can be obtained.
\begin{corollary} \rm
\label{co4-8-11}
For any positive integers $t$ and $q\geq 2$, there always exists an optimal $\overline{t}$-SC$(n,n(q-1),$ $q)$ when $2\leq n< t$.
\end{corollary}

\begin{corollary} \rm
\label{co29-5}
In a $\overline{t}$-SC$(n,M,q)$ with $n=t$ we have
\begin{enumerate}
\item[(i)] $M\leq q^2$, if $n\leq q$,
\item[(ii)] $M\leq nq$, if $n> q$.
\end{enumerate}
\end{corollary}

Using random choice with expurgation, Blackburn \cite{BSC} showed the following result.
\begin{theorem} \rm(\cite{BSC})
\label{BSClower}
Let $n$ and $t$ be fixed integers such that $n \ge 2$ and $t \ge 3$. There exists a positive constant ${\kappa}$, depending only on $n$ and $t$,
so that there is a $q$-ary $\overline{t}$-separable code of length $n$ with at least ${\kappa}q^{n/(t-1)}$ codewords for all sufficiently large integers $q$.
\end{theorem}

\begin{corollary} \rm
\label{coBSC}
There exists a positive constant ${\kappa}$, depending only on $n$ and $t$, so that
$M(\overline{t},n,q) \geq {\kappa}q^{n/(t-1)}$ for all sufficiently large integers $q$.
\end{corollary}

Cheng and Miao \cite{C.M} showed that long-length separable codes can be constructed by concatenating short-length separable codes.
This stimulates the investigation of separable codes with short length $n=2,3$.
When $n=2,3$, upper bounds on $M(\overline{2},n,q)$ were derived,
and several infinite series of optimal $\overline{2}$-SC$(n,M,q)$s were constructed in \cite{C.F,C.J.M}.
When $t\geq 3$, the structure of $\overline{t}$-SCs becomes more complex so that little is known about $\overline{t}$-SCs for $t\geq 3$.

In this paper, we mainly study $\overline{3}$-SCs.
When $n=2$, by Corollary \ref{co4-8-11}, we know that for any integer $q\geq 2$, there always exists an optimal $\overline{3}$-SC$(2,$ $2(q-1),q)$.
When $n=3$ and $q=2$, an exhaustive computer search shows that
 $\mathcal{C}=\{(1,0,0)^T, (0,1,0)^T,(0,0,1)^T\}$ is an optimal $\overline{3}$-SC$(3,3,2)$.
When $q>2$, we have $M(\overline{3},3,q)\leq q^2$ by Corollary \ref{co29-5}. However, this upper bound is not tight.
In the following sections, we will first derive a new upper bound $M(\overline{3},3,q)\leq \lfloor\frac{3q^2}{4}\rfloor$
by considering an optimization problem related to a partial Latin square.
Two constructions for $\overline{3}$-SC$(3,M,q)$s will be then provided by means of perfect hash families and Steiner triple systems.
The first one shows that $\overline{3}$-SC$(3,M,q)$s with $M=O(q^{3/2})$ codewords,
the lower bound provided by Blackburn (Corollary~\ref{coBSC}) using a probabilistic proof, can be constructed explicitly for all integers $q$.
The second one shows that $\overline{3}$-SC$(3,M,q)$s with $M=O(q^2)$ codewords do exist.

\section{Forbidden configurations}%
\label{forb}                                      %

In this section, we first show that a $\overline{3}$-separable code, $\mathcal{C}$, cannot contain certain subcodes,
that is, there are certain forbidden configurations in $\mathcal{C}$.
Then we use these forbidden configurations, together with the $2$-frameproof code property,
to give a necessary and sufficient condition for a $(3,M,q)$ code to be a $\overline{3}$-SC$(3,M,q)$.




For any $(3,M,q)$ code $\mathcal{C}$ defined on $Q$, 
we define two shortened codes $A^{(1)}_i$ and $A^{(1,2)}_{i,k}$ for $i,k \in Q$ as
\begin{eqnarray*}
\label{aij}
A^{(1)}_i= \{(c_2,c_3)\ |\ (i,c_2,c_3)^T\in \mathcal{C}\}\ \hbox{and}\
A^{(1,2)}_{i,k}= \{c_3\  | \ (i, k, c_3)^T \in \mathcal{C}\},
\end{eqnarray*}
respectively. Obviously, $A^{(1)}_i \subseteq Q^2$ and $A^{(1,2)}_{i,k}\subseteq Q$ hold for any $i,k \in Q$,
and $$\sum_{i\in Q}|A^{(1)}_i|=\sum_{i\in Q}\sum_{k\in Q}|A^{(1,2)}_{i,k}|=M.$$
Similarly, $A^{(2)}_i$ and $A^{(3)}_i$ for $i \in Q$ can be defined. Cheng et al. \cite{C.J.M} obtained the following result.

\begin{lemma} \rm{\cite{C.J.M}}
\label{le28-5-1}
If $\mathcal{C}$ is a $\overline{t}$-SC$(n,M,q)$, then $|A_i^{(j)} \bigcap A_{i'}^{(j)}| \le 1$ holds for any distinct $i,i'\in Q$ and $j=1,2,\ldots,n$.
\end{lemma}

From the notions above and the definition of an FPC, the following Lemma \ref{le210} can be obtained.

\begin{lemma} \rm
\label{le210}
A $(3,M,q)$ code is a $2$-FPC$(3,M,q)$ if and only if $|A_{i}^{(j)} \bigcap A_{i'}^{(j)}|\leq 1$ holds
for any $j\in \{1,2,3\}$ and distinct $i,i'\in Q$, where if $|A_{i}^{(j)} \bigcap A_{i'}^{(j)}|= 1$, then $|A_{i}^{(j)}|=|A_{i'}^{(j)}|= 1$.
\end{lemma}

{\bf Proof:} Let $\mathcal{C}$ be a $(3,M,q)$ code. We first suppose that $\mathcal{C}$ is a $2$-FPC$(3,M,q)$.
According to Lemma~\ref{le28-5}, $\mathcal{C}$ is a $\overline{2}$-SC$(3,M,q)$.
So for any $j \in \{1,2,3\}$ and distinct $i,i' \in Q$, $|A_i^{(j)} \bigcap A_{i'}^{(j)}| \le 1$ holds by Lemma~\ref{le28-5-1}.
When $|A_{i}^{(j)} \bigcap A_{i'}^{(j)}|= 1$, without loss of generality, we may assume that $j=1$ and $\{ {\bf a}\}=A_{i}^{(1)} \bigcap A_{i'}^{(1)}$.
If there is another element ${\bf b}\in A_{i}^{(1)}$, then we have $(i,{\bf a})^T$, $(i,{\bf b})^T$, $(i',{\bf a})^T \in \mathcal{C}$,
$\{(i',{\bf a})^T, (i,{\bf b})^T\} \bigcap \{(i,{\bf a})^T\} =\emptyset$,  and $(i,{\bf a})^T \in {\sf desc}(\{(i',{\bf a})^T, (i,{\bf b})^T\}) \bigcap \mathcal{C}$.
This contradicts the definition of a frameprood code. So $|A_{i}^{(1)}|= 1$ holds. Similarly, $|A_{i'}^{(1)}|=1$ can be proved.

Conversely, if $\mathcal{C}$ is not a $2$-FPC$(3,M,q)$, then there exist three distinct codewords
${\bf a} = (a_1,a_2$, $a_3)^T$, ${\bf b} = (b_1,b_2,b_3)^T$, ${\bf c} = (c_1,c_2,c_3)^T \in \mathcal{C}$,
such that ${\bf c} \in {\sf desc}(\{{\bf a}, {\bf b}\})$.
Since the codeword length equals $3$, there exists at least one codeword, say ${\bf a}$,
such that it contains two same coordinates as ${\bf c}$, say $a_2=c_2$ and $a_3=c_3$.
Since ${\bf c} \in {\sf desc}(\{{\bf a}, {\bf b}\})$, then $b_1=c_1$.
That is, 
${\bf c}=(b_1,a_2,a_3)^T$.
Obviously, $|A_{a_{1}}^{(1)} \bigcap A_{b_{1}}^{(1)}|\geq 1$ and $|A_{b_{1}}^{(1)}|\geq 2$, which contradict our assumption.
\qed

Now, let us turn our attention to $\overline{3}$-SC$(3,M,q)$s. For any $\overline{3}$-SC$(3,M,q)$, $\mathcal{C}$,
it can be checked that there is no subcode $\triangle_{1}\subseteq \mathcal{C}$ described in (\ref{fo12}).
 \begin{eqnarray}
\label{fo12}
\triangle_{1}=\left(
                     \begin{array}{cccc}
                       a & a & b & b \\
                       e & f & g & e \\
                       c & d & c & d
                     \end{array}
                   \right), \ \ \
\triangle_{2}=\left(
                     \begin{array}{cccc}
                       a & a & b & b \\
                       c & d & c & d \\
                       e & f & g & e
                     \end{array}
                   \right), \ \ \
\triangle_{3}=\left(
                     \begin{array}{cccc}
                       e & f & g & e \\
                       a & a & b & b \\
                       c & d & c & d
                     \end{array}
                   \right).
\end{eqnarray}
The reason is as follows. If $\mathcal{C}$ contains the configuration $\triangle_{1}$,
then since $\mathcal{C}$ is also a $2$-FPC$(3,M,q)$ by Theorem \ref{adle1}, we should have $a\neq b$, $c\neq d$ and $e\not\in\{f,g\}$.
However, in this case, $\{(a,e,c)^T$, $(a,f,d)^T,(b,g,c)^T\} \neq \{(a,f,d)^T,(b,g,c)^T,(b,e,d)^T\}$,
but ${\sf desc}(\{(a,e,c)^T,(a,f,d)^T,(b,g,c)^T\})={\sf desc}(\{(a,f,d)^T,(b,g,c)^T,(b,e,d)^T\})$, a contradiction to the definition of a $\overline{3}$-SC.
We call such $\triangle_{1}$ a {\em forbidden configuration} of $\mathcal{C}$.
Given a subcode $\mathcal{C}' \subseteq \mathcal{C}$, conjugates of $\mathcal{C}'$ are subsets of $Q^3$ defined by
changing any two coordinates of $\mathcal{C}'$.
Clearly, $\triangle_{1}$ and its conjugates in (\ref{fo12}) are forbidden configurations of $\mathcal{C}$.

It is easy to check that the following $\nabla$ is also a forbidden configuration of $\mathcal{C}$, where $|\{a_i,b_i,c_i\}|=3$, $i=1,2,3$.
\begin{eqnarray}
\label{fo13}
\nabla=\left(
                   \begin{array}{cccccc}
                    a_{1} & b_{1} & c_{1} &  a_{1} & b_{1} & c_{1} \\
                    a_{2} & b_{2} & c_{2} &  b_{2} & c_{2} & a_{2} \\
                    a_{3} & b_{3} & c_{3} &  c_{3} & a_{3}&  b_{3}
                   \end{array}
                 \right)\ \ \ \
\end{eqnarray}

\begin{theorem} \rm
\label{th2610}
A $(3,M,q)$ code $\mathcal{C}$ is a $\overline{3}$-SC$(3,M,q)$ if and only if it satisfies the following conditions:
\begin{enumerate}
\item[(i)] $\mathcal{C}$ is a $2$-FPC$(3,M,q)$;
\item[(ii)] Configurations in (\ref{fo12}) and (\ref{fo13}) are all the forbidden configurations of $\mathcal{C}$.
\end{enumerate}
\end{theorem}
The proof of Theorem~\ref{th2610} is included in Appendix.

\section{An upper bound}    %
\label{upper}                        %

As we said at the end of Section \ref{pre}, when $n=3$ and $q>2$, we have an upper bound $M(\overline{3},3,q)\leq q^2$.
In this section, we are going to derive a new upper bound $M(\overline{3},3,q)\leq\lfloor\frac{3q^2}{4}\rfloor$
by exploiting the two conditions in Theorem \ref{th2610}.

Given a $\overline{3}$-SC$(3,M,q)$, $\mathcal{C}$, defined on $Q=\{0,1,\ldots,q-1\}$,
the following $q\times q$ array can be obtained, where each entry is a subset of $Q$.
\begin{eqnarray*}
\label{eq21023}
\mathfrak{A}= \left(\begin{array}{cccccccc}
{\mathcal{A}}_{0,0}^{(1,2)} & \ldots  & {\mathcal{A}}_{0,q-1}^{(1,2)} \\[0.3cm]
{\mathcal{A}}_{1,0}^{(1,2)} & \ldots  & {\mathcal{A}}_{1,q-1}^{(1,2)} \\[0.3cm]
\ldots &  \ldots & \ldots  \\[0.3cm]
{\mathcal{A}}_{q-1,0}^{(1,2)} & \ldots & {\mathcal{A}}_{q-1,q-1}^{(1,2)} \\
\end{array}\right)
\end{eqnarray*}

By Theorem \ref{th2610}, $\mathcal{C}$ is also a $2$-FPC$(3,M,q)$.
Then it can be easily checked that $\mathfrak{A}$ has the following properties.\\[0.2cm]
(I) P$_{1.1}$: If $|{\mathcal{A}}_{i,j}^{(1,2)}| \ge 2$, then
${\mathcal{A}}_{i,j}^{(1,2)}\bigcap {\mathcal{A}}_{i',j'}^{(1,2)}=\emptyset$ holds for any $(i',j')\in Q^2\setminus \{(i,j)\}$;

 P$_{1.2}$: If $|{\mathcal{A}}_{i,j}^{(1,2)}\bigcap{\mathcal{A}}_{i,j'}^{(1,2)}|= 1$ with $j\neq j'$, then
${\mathcal{A}}_{i',j}^{(1,2)}={\mathcal{A}}_{i',j'}^{(1,2)}=\emptyset$ holds for any $i'\in Q\setminus\{i\} $;

  P$_{1.3}$: If $|{\mathcal{A}}_{i,j}^{(1,2)}\bigcap{\mathcal{A}}_{i',j}^{(1,2)}|= 1$ with $i\neq i'$, then
  ${\mathcal{A}}_{i,j'}^{(1,2)}={\mathcal{A}}_{i',j'}^{(1,2)}=\emptyset$ holds for any $j'\in Q\setminus\{j\} $.

\vskip6pt

Again by Theorem \ref{th2610}, configuration $\triangle_2$ is a forbidden configuration of $\mathcal{C}$.
Then it is easily seen that $\mathfrak{A}$ has the following property.\\[0.2cm]
(II) P$_{2}$: If $|{\mathcal{A}}_{i,j}^{(1,2)}\bigcap{\mathcal{A}}_{i',j'}^{(1,2)}|=1$ with $i\neq i'$, $j\neq j'$,
then ${\mathcal{A}}_{i',j}^{(1,2)}=\emptyset$ or ${\mathcal{A}}_{i,j'}^{(1,2)}=\emptyset$ holds.

\vskip6pt

From array $\mathfrak{A}$, we can define a related array ${\mathbb{B} }=(b_{i,j})$ in the following way:
\begin{eqnarray*}
b_{i,j}=\left\{\begin{array}{ll}
a_{i,j} & \  \mbox{if}\ \ {\mathcal{A}}^{(1,2)}_{i,j}=\{a_{i,j}\},\\
\star & \  \mbox{if}\ \ 2 \le |{\mathcal{A}}^{(1,2)}_{i,j}|,\\
\times & \  \mbox{if } {\mathcal{A}}^{(1,2)}_{i,j}=\emptyset.
\end{array}
\right.
\end{eqnarray*}
Let $\mu_{i}=|\{j\ |\ |{\mathcal{A}}_{i,j}^{(1,2)}|\geq 2, j\in Q\}|$ be the number of sets ${\mathcal{A}}_{i,j}^{(1,2)}$
with $|{\mathcal{A}}_{i,j}^{(1,2)}| \ge 2$ in the $i$th row of $\mathfrak{A}$, namely, the number of ``$\star$" in the $i$th row of ${\mathbb{B} }$.
Let $\rho_{i}=\sum_{j\in Q,2\leq |{\mathcal{A}}_{i,j}^{(1,2)}|}|{\mathcal{A}}_{i,j}^{(1,2)}|$ be the sum of orders of
${\mathcal{A}}_{i,j}^{(1,2)}$ with $|{\mathcal{A}}_{i,j}^{(1,2)}| \ge 2$ in the $i$th row of $\mathfrak{A}$.
Then according to P$_{1.1}$, we know that the sum of orders of ${\mathcal{A}}_{i,j}^{(1,2)}$
with $|{\mathcal{A}}_{i,j}^{(1,2)}| \ge 2$ in $\mathfrak{A}$ is
\begin{eqnarray*}
\sum_{i\in Q}\rho_i=|\bigcup\limits_{ \substack{i,j\in Q \\ |{\mathcal{A}}_{i,j}^{(1,2)}|\geq 2 }}{\mathcal{A}}_{i,j}^{(1,2)}|
=\sum\limits_{ \substack{i,j\in Q\\ |{\mathcal{A}}_{i,j}^{(1,2)}|\geq 2 }}|{\mathcal{A}}_{i,j}^{(1,2)}|.
\end{eqnarray*}

Let $D_i$ be the set of elements of $Q$ occurring at least twice in the $i$th row of ${\mathbb{B}}$.
Let $m_i$ be the sum of frequencies of elements in $D_i$ in the $i$th row of $\mathbb{B}$,
$m(-i)=\sum_{l\in Q \setminus \{i\}}m_{l}$ and $m=\sum_{l\in Q}m_{l}$.
%
%
Let $B_i$ be the set of elements of $Q$ occurring exactly once in the $i$th row of ${\mathbb{B}}$. Let $z_{i,j}=|B_{i}\bigcap B_{j}|$.
Clearly, $z_{i,j}\leq q$.

Now we consider the total number of distinct elements of $Q$ in the $i$th and $j$th rows of $\mathbb{B}$.
The number of distinct elements of $Q$ in any $l$th row of $\mathbb{B}$ is $|B_l|+|D_l|$.
The number of distinct elements of $Q$ in the $i$th and $j$th rows of $\mathbb{B}$ is
at least $|B_i|+|B_j|-|B_i \cap B_j|$, and from the definition of $\rho_l$ and property P$_{1,1}$, is at most $q-\sum_{i\in Q}\rho_i$. So we have
$$|B_i|+|B_j|-|B_i \cap B_j| \leq q-\sum_{l\in Q}\rho_{l}.$$

We also consider the number of entries in the $i$th and $j$th rows of $\mathbb{B}$.
Clearly, for the $l$th row of $\mathbb{B}$, $m(-l)+\mu_l+m_l+|B_l|\leq q$, and for the $i$th and $j$th rows of $\mathbb{B}$,
$$[|B_i|+\mu_{i}+m(-i)+m_i]+[|B_j|+\mu_{j}+m(-j)+m_j]+z_{i,j}\leq 2q$$
where the term ``$z_{i,j}$" is the number of additional ``$\times$" caused by the forbidden configuration $\triangle_2$. 

We are in a position to derive our new upper bound on $M(\overline{3},3,q)$.
\begin{theorem} \rm
\label{th2617}
$M(\overline{3},3,q) \le \lfloor\frac{3q^2}{4}\rfloor$ holds for $q\geq 4$.
\end{theorem}
{\bf Proof:} Let $\mathcal{C}$ be a $\overline{3}$-SC$(3,M,q)$ defined on $Q$.
Clearly, $M=\sum_{i,l\in Q}|{\mathcal{A}}_{i,l}^{(1,2)}|$, where $\sum_{l\in Q}|{\mathcal{A}}_{i,l}^{(1,2)}|=\rho_i+$ $m_i+|B_i|$.
So our goal is to solve the following optimization problem.
\begin{eqnarray*}
\label{eq211}
\left\{ \begin{array}{llll} \hbox{Maximize} &  \sum_{i\in Q}(\rho_i+m_i+|B_i|)\\[0.4cm]
  \hbox{Subject to}   &2m+\mu_i+|B_i|+\mu_j+|B_j|+z_{i,j}\leq 2q\\[0.2cm]
  & |B_i|+|B_j|-z_{i,j}+\sum_{l\in Q}\rho_{l}\leq q\\[0.2cm]
 &  i, j\in Q.
\end{array} \right.
\end{eqnarray*}
From the constraints above, we have
\begin{eqnarray*}
\label{eq211}
2m+\mu_i+\mu_j+2(|B_i|+|B_j|)+\sum_{l\in Q}\rho_{l}\leq 3q.
\end{eqnarray*}
Summarizing the inequality above for all $i\neq j$, we have
\begin{eqnarray*}
&&2(q-1)\sum_{l\in Q}|B_l|+(q-1)\sum_{l\in Q}\mu_l+2m{q\choose 2}+{q\choose 2}\sum_{l\in Q}\rho_{l}\leq 3q{q\choose 2}.
\end{eqnarray*}
When $2(q-1)\leq {q\choose 2}$, that is, $q\geq 4$, we have
\begin{eqnarray*}
\begin{split}
2(q-1)M&=2(q-1)\sum_{l\in Q}(|B_l|+m_l+\rho_{l})\\
& \leq 2(q-1)\sum_{l\in Q}(|B_l|+m_l)+{q\choose 2}\sum_{l\in Q}\rho_{l}\\
&\leq 3q{q\choose 2}-(q-1)\Bigr(\sum_{l\in Q}\mu_{l}+(q-2)m\Bigl)\\
&\leq 3q{q\choose 2}.
\end{split}\end{eqnarray*}
This implies $M\leq \lfloor\frac{3q^2}{4}\rfloor$.\qed

This bound is sharp for some $q\geq 4$. In fact, if we can find a $\overline{3}$-SC$(3,M,q)$, $\mathcal{C}$, on $Q$
such that every entry in its corresponding $\mathfrak{A}$ is a singleton or empty set,
no element of $Q$ appears in any row or column of $\mathfrak{A}$ more than once,
and the number of ``$\times$" in each row of $\mathbb{B}$ is $\frac{q}{4}$, then $M=\frac{3}{4}q^2$.
The following is such an example.
\begin{exmp} \rm
\label{ex12}
When $q=4$, $M(\overline{3},3,4)= 3\times4^2/4=12$.
Then it can be checked that $\mathcal{C}_4$ is an optimal $\overline{3}$-SC$(3,12,4)$.
We list it and its related array as follows.
\begin{eqnarray*}
\mathcal{C}_4=\left(\begin{array}{cccccccccccccccc}
0 & 0 & 0 & 1 & 1 & 1 & 2 & 2 & 2 & 3 & 3 & 3\\
0 & 1 & 2 & 0 & 2 & 3 & 0 & 1 & 3 & 1 & 2 & 3 \\
0 & 1 & 2 & 1 & 3 & 2 & 3 & 2 & 0 & 3 & 0 & 1
\end{array}\right),\ \
\mathbb{B}_4=\left(\begin{array}{cccccccccccccccc}
0      & 1      & 2      & \times\\
1      & \times & 3      & 2 \\
3      & 2      & \times & 0\\
\times & 3      & 0      & 1
\end{array}\right)
\end{eqnarray*}
\end{exmp}

\section{Constructions} %
\label{construction}       %

$\mathbb{B}_4$ in Example \ref{ex12} can be regarded as a partial Latin square of order $4$,
which corresponds to an optimal  $\overline{3}$-SC$(3,12,4)$.
In this section, we use partial Latin squares to construct $\overline{3}$-SC$(3,M,q)$s.
The definition of a partial Latin square is given below.



\begin{definition} \rm
\label{d12}
A {\em partial Latin square} $\mathbb{P}$ of order $q$ is a $q\times q$ array with entries chosen from a $q$-set $Q$
in such a way that each element of $Q$ occurs at most once in each row and at most once in each column of the array.
If all the cells of the array are filled then the partial Latin square is termed a Latin square.
\end{definition}

For ease of exposition, a partial Latin square $\mathbb{P}$ will be represented by a set of ordered triples
$\{(i,j,P_{ij})$ $|$ element $P_{ij}$ occurs in cell $(i,j)$ of the array$\}$.


\begin{lemma} \rm
\label{le1319}
For any partial Latin square $\mathbb{P}$, its associated code
$\mathcal{C} = \{{\bf c} \ | \ {\bf c} = (i,j,P_{ij})^{T} \ | $ $(i,j;P_{ij}) \in \mathbb{P}\}$
is a $2$-FPC.
\end{lemma}
{\bf Proof:} Suppose $\mathcal{C}$ is not a $2$-FPC. Then there exist $\mathcal{C}' = \{{\bf a}, {\bf b}\} \subseteq \mathcal{C}$
and ${\bf c}=(c_1,c_2,c_3)^{T} \in \mathcal{C}\setminus \mathcal{C}'$ such that ${\bf c}\in {\sf desc}(\mathcal{C}')$.
This implies that there exists a codeword, say ${\bf b}=(b_1,b_2,b_3)^T \in \mathcal{C}'$,
such that there are at least two coordinates $i,j\in \{1,2,3\}$ satisfying $b_i=c_i$ and $b_j=c_j$.
If $(i,j)=(1,2)$, then $\{(c_1,c_2;c_3),(c_1,c_2;b_3)\} \subseteq\mathbb{P}$, that is, cell $(c_1,c_2)$ contains both $c_3$ and $b_3$,
a contradiction to the definition of a partial Latin square. Similarly, it is readily checked that neither $(i,j)=(1,3)$ nor $(2,3)$ is possible.
The proof is then completed.
\qed

\vskip6pt

In the following subsections, we first construct partial Latin squares by means of perfect hash families and Steiner triple systems, respectively,
and then show that the associated codes of these partial Latin squares do not contain the forbidden configurations (\ref{fo12}) and (\ref{fo13}).
According to Theorem \ref{th2610}, these associated codes are $\overline{3}$-SC$(3,M,q)$s.


\subsection{Constructions via perfect hash families}

Let $f$ be a function from a set $X$ to a set $Y$.
We say that $f$ separates a subset $T\subseteq X$ if $f$ is injective when $f$ is restricted to $T$;
otherwise we say that $f$ reduces $T$. Let $M, q, t$ be integers such that $M \geq q \geq t \geq 2$.
Suppose $|X|=M$ and $|Y|=q$.
A set $\mathcal{F}$ of functions from $X$ to $Y$ with $|\mathcal{F}|=n$ is an $(n;M,q,t)$-{\em perfect hash family}
if for all $T\subseteq X$ with $|T| = t$, there exists at least one $f\in \mathcal{F}$ such that $f$ separates $T$.
We use the notation PHF$(n;M,q,t)$ for an $(n;M,q,t)$-perfect hash family.
A PHF$(n;M,q,t)$ can be depicted as an $n \times M$ array in which the columns are labeled by the elements $j \in X$,
the rows by the functions $f_i\in \mathcal{F}$, and the $(i, j)$-entry of the array is the value $f_i(j)$.
Thus, a PHF$(n;M,q,t)$ is equivalent to an $n \times M$ array with entries from a set of $q$ symbols
such that every $n \times t$ subarray contains at least one row having distinct symbols.
A perfect hash family is optimal if $M$ is the largest possible value given $n$, $q$ and $t$.

Given a PHF$(n;M,q,t)$, $\mathcal{F} = \{f_1, f_2, \ldots, f_n\}$, we can derive an associated code $\mathcal{C}$ in an obvious way:
associate each $x \in X$ with the codeword $(f_1(x), f_2(x), \ldots, f_n(x))^T$.
Staddon et al. \cite{S.S.W} observed that the associated code of a PHF$(n;M,q,t)$ is a $(t-1)$-FPC$(n,M,q)$.
However, we should note that the associated codes of a PHF$(n;M,q,t)$ is not always a $\overline{t}$-SC$(n,M,q)$.
In fact, we can easily check that $\nabla$ in (\ref{fo13}), a forbidden configuration of $\overline{3}$-SC$(3,M,q)$, is a PHF$(3;6,q,3)$.

In this subsection, we will first show that a special optimal PHF$(3;r^3,r^2,3)$ in \cite{B4} is in fact also a $\overline{3}$-SC$(3,r^3,r^2)$.
Based on this PHF$(3;r^3,r^2,3)$, we then propose a new construction for a $\overline{3}$-SC$(3,r^3+r^{5/2},r^2)$ for any even square $r$.

Let $r\geq 2$ be an integer, $X=Z_r^3$, and $Y=Z_{r^2}$. Define functions $f_1$, $f_2$, $f_3$ : $X\rightarrow Y$ by
\[  f_1((a, b, c))=ar+b,\ \
f_2((a, b, c))=ar+c\ \  \hbox{and}\ \
f_3((a, b, c))=br+c\]
for all $a, b, c \in Z_r$.

\begin{theorem} \rm(\cite{B4})
\label{th4-12} The functions $f_1$, $f_2$, and $f_3$ defined above form an optimal PHF$(3;r^3,r^2,3)$.
\end{theorem}

Let $\mathcal{C}_1 = \{(f_1((a,b,c)),f_2((a,b,c)),f_3((a,b,c)))^T \ |\ a,b,c\in Z_r\}$ be the associated code of the above PHF.
It is not difficult to check that the corresponding $r^2 \times r^2$ array of $\mathcal{C}_1$, in which
the rows are labeled by $f_1((a,b,c))$, the columns by $f_2((a,b,c))$, and $(f_1((a,b,c)),f_2((a,b,c)))$-entry is $f_3((a,b,c))$, is a partial Latin square.
Then Lemma \ref{le1319} shows that $\mathcal{C}_1$ is a $2$-FPC$(3,r^3,r^2)$.
In fact, we can say more about $\mathcal{C}_1$.

\begin{lemma} \rm
\label{lemma421-12} $\mathcal{C}_1$ is a $\overline{3}$-SC$(3,r^3,r^2)$ for any integer $r \ge 2$.
\end{lemma}
{\bf Proof:} From Theorem \ref{th2610}, it is sufficient to prove that
forbidden configurations in (\ref{fo12}) and (\ref{fo13}) cannot occur in $\mathcal{C}_1$.
It is easy to check that the forbidden configurations in (\ref{fo12}) are also forbidden configurations of PHF$(3;r^3,r^2,3)$.
Now, let us consider the forbidden configuration in (\ref{fo13}).
Denote a $6$-subset of $\mathcal{C}_1$ by $C=(\mbox{\boldmath$\alpha$}_1,\mbox{\boldmath$\alpha$}_2,\ldots,\mbox{\boldmath$\alpha$}_6)$,
where $\mbox{\boldmath$\alpha$}_{i}=(a_{i}r+b_{i},a_{i}r+c_{i},b_{i}r+c_{i})^T$, $1\leq i\leq 6$.
Assume $C=\nabla$, then we have
\begin{eqnarray*}
\begin{array}{lll}
a_1r+b_1=a_4r+b_4\ \ \ &
a_2r+b_2=a_5r+b_5\ \ \ &
a_3r+b_3=a_6r+b_6\\
a_1r+c_1=a_6r+c_6\ \ \ &
a_2r+c_2=a_4r+c_4\ \ \ &
a_3r+c_3=a_5r+c_5\\
b_1r+c_1=b_5r+c_5\ \ \ &
b_2r+c_2=b_6r+c_6\ \ \ &
b_3r+c_3=b_4r+c_4
\end{array}.
\end{eqnarray*}
which imply $a_1=a_2$, $b_1=b_2$ and $c_1=c_2$, a contradiction to the assumption. Then the proof is completed.\qed

From Lemma \ref{lemma421-12}, a lower bound can be obtained.

\begin{corollary}\rm
\label{co29-7} $\lfloor\sqrt{q}\rfloor^3 \leq M(\overline{3},3,q)$ holds for $q \ge 4$.
\end{corollary}

Now we illustrate Lemma \ref{lemma421-12} with a small example.

\begin{exmp} \rm
\label{ex1}
(1) Let $r=4$. Then according to Lemma \ref{lemma421-12}, the associated code of $\mathbb{B}_1$ described below is a $\overline{3}$-SC$(3,64,16)$.
\begin{eqnarray*}
\mathbb{B}_1=\left(\begin{array}{cccc}
A &    &    &  \\
   & A &    &  \\
   &    & A &  \\
   &    &    &  A
\end{array}\right),\ \ \hbox{where}\
A=\left(\begin{array}{ccccc}
0&1&2&3&\\
4&5&6&7&\\
8&9&10&11&\\
12&13&14&15
\end{array}\right)
\end{eqnarray*}
and other cells of $\mathbb{B}_1$ are filled with ``$\times$".

(2) Replacing some ``$\times$"s in ${\mathbb B}_1$ by some elements of $Y=Z_{r^2}$, we obtain the following array ${\mathbb B}$,
which is still a partial Latin square. It can be checked that its associated code is a $\overline{3}$-SC$(3,96,16)$.
\begin{eqnarray*}
\label{eq38}
\mathbb{B}=\left(\begin{array}{cccc}
A  & A_1 &    &  \\
   & A   & A_2&  \\
   &     & A  & A_1\\
A_2&     &    & A
\end{array}\right),\ \ \hbox{where}\
A_1=\left(\begin{array}{ccccc}
10&11& & \\
14&15& & \\
 & &0&1\\
 & &4&5
\end{array}\right),\ \
A_2=\left(\begin{array}{ccccc}
 & &8&9\\
 & &12&13\\
2&3& &  \\
6&7& &
\end{array}\right).
\end{eqnarray*}
\end{exmp}

Inspired by Example \ref{ex1}, for any even square $r=k^2$, we define other six functions as follows:
\begin{eqnarray*}
\label{fun1}
\begin{array}{ll}g_1(x,y,z,h)=xk+y+hr, &g_2(x,y,z,h)=xk+z+(h+1)r,\ \ \ \ \ \ \\
g(x,y)=r-(x+1)k+y,& g_3(x,y,z,h)=g(x,y)r +g(x,z),\\
g'_2(x,y,z,h)=g(x,z)+(h+1)r,\ &g'_3(x,y,z,h)=g(x,y)r +xk+z,
\end{array}\end{eqnarray*}
where all the operations in the above functions are taken in $Z_{r^2}$. Using these functions, we further define the following codes.
\begin{eqnarray*}
\mathcal{C}_2&=&\{(g_1(x,y,z,h),g_2(x,y,z,h),g_3(x,y,z,h))^T \ |\ x,y,z\in Z_k, h\in Z_r, h\equiv\ 0\ mod\ 2 \}, \\
\mathcal{C}_3&=&\{(g_1(x,y,z,h), g'_2(x,y,z,h),g'_3(x,y,z,h))^T\ |\ x,y,z\in Z_k, h\in Z_r, h\equiv\ 1\ mod\ 2 \}.
\end{eqnarray*}

Similar to Lemma \ref{lemma421-12}, we have
\begin{lemma}\rm
\label{lec2-c3}
For any even square $r=k^2$, $\mathcal{C}_2$ is a $\overline{3}$-SC$(3,kr^2/2,r^2)$, and so is $\mathcal{C}_3$.
\end{lemma}
{\bf Proof:} Let us first consider $\mathcal{C}_2$. Let $X=Z_k^3 \times Z_r$ and $Y=Z_{r^2}$.
We show that $\mathcal{F}' = \{g_1,g_2,g_3\}$ forms a PHF$(3;kr^2/2, r^2,3)$.
It is easy to check that an element ${\bf v}\in X$ is uniquely determined by any two of three images $g_1({\bf v})$, $g_2({\bf v})$ and $g_3({\bf v})$,
therefore every $2$-subset of $X$ is reduced by at most one of $g_1$, $g_2$ and $g_3$.
Suppose, for a contradiction, that $\{{\bf v}_1=(x_1,y_1,z_1,h_1), {\bf v}_2=(x_2,y_2,z_2,h_2), {\bf v}_3=(x_3,y_3,z_3,h_3)\} \subseteq X$
is a $3$-set that is reduced by all of the functions $g_1$, $g_2$ and $g_3$.
Since every $2$-set is reduced by at most one of $g_1$, $g_2$ and $g_3$ and since every function $g_i$ must reduce
some $2$-subset of $\{{\bf v}_1, {\bf v}_2, {\bf v}_3\}$, we may assume without loss of generality that
$g_1({\bf v}_1)=g_1({\bf v}_2)$, $g_2({\bf v}_1)=g_2({\bf v}_3)$ and $g_3({\bf v}_2)=g_3({\bf v}_3)$.
This implies that $x_1 =x_2$, $y_1 =y_2$, $z_1 =z_2$ and $h_1 =h_2$, a contradiction to the assumption.
So $\mathcal{F}'$ is a $3$-PHF$(3;kr^2/2,r^2,3)$.

Clearly, $\mathcal{C}_2$ is the associated code of the perfect hash family $\mathcal{F}'$.
Similar to the proof of Lemma \ref{lemma421-12}, we need only consider the forbidden configuration in (\ref{fo13}).
Denote a $6$-subset of $\mathcal{C}_2$ by $C=(\mbox{\boldmath$\beta$}_1,\mbox{\boldmath$\beta$}_2,\ldots,\mbox{\boldmath$\beta$}_6)$,
where $\mbox{\boldmath$\beta$}_{i}=(g_1(x_i,y_i,z_i,h_i),g_2(x_i,y_i,z_i,h_i),g_3(x_i,y_i,z_i,h_i))^T$, $1\leq i\leq 6$.
Assume $C=\nabla$, then we have
\begin{small}
\begin{eqnarray*}
&& x_1k+y_1+h_{1}r = x_4k+y_4+h_{4}r, \ \ \ \ \ \ \ \ \ \ \ \ \ \ \ \ \   x_2k+y_2+h_{2}r = x_5k+y_5+h_{5}r, \\
&& x_3k+y_3+h_{3}r = x_6k+y_6+h_{6}r, \ \ \ \ \ \ \ \ \ \ \ \ \ \ \ \ \   x_1k+z_1+(h_1+1)r = x_6k+z_6+(h_6+1)r, \\
&& x_2k+z_2+(h_2+1)r = x_4k+z_4+(h_4+1)r, \ \ \                   x_3k+z_3+(h_3+1)r = x_5k+z_5+(h_5+1)r, \\
&& r[r-(x_{1}+1)k+y_{1}]+r-(x_1+1)k+z_1 = r[r-(x_{5}+1)k+y_{5}]+r-(x_5+1)k+z_5, \\
&& r[r-(x_{2}+1)k+y_{2}]+r-(x_2+1)k+z_2 = r[r-(x_{6}+1)k+y_{6}]+r-(x_6+1)k+z_6, \\
&& r[r-(x_{3}+1)k+y_{3}]+r-(x_3+1)k+z_3 = r[r-(x_{4}+1)k+y_{4}]+r-(x_4+1)k+z_4.
\end{eqnarray*}
\end{small}
\noindent That is, $x_3=x_6$, $y_3=y_6$, $z_3=z_6$ and $h_3=h_6$, a contradiction to the assumption.
By Theorem \ref{th2610}, we know that $\mathcal{C}_2$ is a $\overline{3}$-SC$(3,kr^2/2,r^2)$.

In a similar fashion, we can prove that $\mathcal{C}_3$ is also a $\overline{3}$-SC$(3,kr^2/2,r^2)$. The proof is then completed. \qed

As a matter of fact, $A_1$ in Example \ref{ex1} corresponds to $\mathcal{C}_2$ with $k=2$,
while $A_2$ corresponds to $\mathcal{C}_3$ with $k=2$.

From the constructions of $\mathcal{C}_i$, $i=1,2,3$, it can be easily checked that the following assertions hold
\begin{corollary}\rm
\label{co6-8}
\begin{enumerate}
\item[(i)] $\mathcal{C}_i\cap\mathcal{C}_j=\emptyset$ for any distinct integers $i,j\in \{1,2,3\}$;
\item[(ii)] The related array of $\mathcal{C} = \mathcal{C}_1\cup\mathcal{C}_2\cup \mathcal{C}_3$ is a partial Latin square;
\item[(iii)] For any $(a,b,c)^T\in \mathcal{C}_2$ and $(d,e,f)^T\in \mathcal{C}_3$, we have $a\neq d$, $b\neq e$ and $c\neq f$.
\end{enumerate}
\end{corollary}


From Corollary \ref{co6-8}, we have the following result.
\begin{theorem}\rm
\label{thc2}
Let $r=k^2$, where $k$ is an even positive integer.
Then $\mathcal{C}=\mathcal{C}_1\cup\mathcal{C}_2\cup \mathcal{C}_3$ is a $\overline{3}$-SC$(3,r^3+kr^2,r^2)$.
\end{theorem}
{\bf Proof:} According to Corollary \ref{co6-8} and Lemma \ref{le1319}, the related array of $\mathcal{C}$ is a partial Latin square,
and $\mathcal{C}$ is a $2$-FPC. Let

\indent\ \ \ \ \ $\mbox{\boldmath$\alpha$}_i=(a_ir+b_i,a_ir+c_i,b_ir+c_i)^T\in \mathcal{C}_1$,\\[0.1cm]
\indent\ \ \ \ \ $\mbox{\boldmath$\beta$}_i=(g_1(x_i,y_i,z_i,h_i),g_2(x_i,y_i,z_i,h_i),g_3(x_i,y_i,z_i,h_i))^T\in \mathcal{C}_2$ \\[0.1cm]
\indent\ \ \ \ \ $\mbox{\boldmath$\gamma$}_i=(g_1(x_i,y_i,z_i,h_i), g'_2(x_i,y_i,z_i,h_i),g'_3(x_i,y_i,z_i,h_i))^T\in \mathcal{C}_3$,

\noindent where $1\leq i\leq 6$. From Theorem \ref{th2610}, to prove that $\mathcal{C}$ is a $\overline{3}$-SC,
we only need consider the forbidden configurations in (\ref{fo12}) and (\ref{fo13}).
\begin{itemize}
\item[(I)] Assume that $\triangle_1=({\bf c}_1, {\bf c}_2,{\bf c}_3,{\bf c}_4)$ occurs in $\mathcal{C}$,
where ${\bf c}_i\in \mathcal{C}_{k_i}$ and $k_i\in\{1,2,3\}$, $i\in\{1,2,3,4\}$.
Let us consider the $4$-tuple vector $(k_1,k_2,k_3,k_4)$.
\begin{itemize}
\item[(1)] Suppose that $(k_1,k_2,k_3,k_4)\in K_1=\{(k,k,k,k)\ |\ k\in\{1,2,3\}\}$.
This implies that the codewords ${\bf c}_1$, ${\bf c}_2$, ${\bf c}_3$, ${\bf c}_4$ are all in the same subcode $\mathcal{C}_k$, $k\in\{1,2,3\}$,
a contradiction to the fact that $\mathcal{C}_k$ is a $\overline{3}$-SC.
\item[(2)] Suppose that $(k_1,k_2,k_3,k_4)\in K_2$, where
$K_2 = \{(2, 3, k_3, k_4)$, $(2, k_2, 3, k_4)$, $(2, k_2, k_3, 3)$, $(k_1, 2, k_3, 3)$, $(k_1, k_2, 2, 3)$, $
(3, 2, k_3, k_4)$, $(3, k_2, 2, k_4)$, $(3, k_2, k_3, 2)$, $(k_1, 3, k_3, 2)$, $(k_1, k_2, 3, 2) \ | \ k_1, k_2, k_3, k_4 \in \{1,2,3\}\}$.
This contradicts the statement (iii) of Corollary \ref{co6-8}.
\item[(3)] Suppose that $(k_1,k_2,k_3,k_4)\in K_3=\{1,2,3\}^4\setminus(K_1\cup K_2)$.
It is easy to check that none of the above cases equals $\triangle_{1}$. We take $(k_1,k_2,k_3,k_4)=(1,1,1,2)$ as an example.
Suppose that $\triangle_{1}=(\mbox{\boldmath$\alpha$}_1$, $\mbox{\boldmath$\alpha$}_2$,
$\mbox{\boldmath$\alpha$}_3$, $ \mbox{\boldmath$\beta$}_4)$. We have\\[0.2cm]
\indent\ \ \ \ \ \  $a_1r+b_1 = a_2r + b_2$, $a_1r+c_1 = x_4k+z_4+(h_4+1)r$, $b_1r+c_1 = b_3r+c_3$,
$b_2r+c_2 = [r-(x_4+1)k+y_4]r+[r-(x_4+1)k+z_4]$, $a_3r+b_3 = x_4k+y_4+h_4r$.\\[0.2cm]
That is, $x_4k + y_4 = b_3 = b_1 = b_2 = r - (x_4 + 1)k + y_4$. This implies $2x_4+1=k$, a contradiction to the fact that $k$ is even.
\end{itemize}
According to (1)-(3) above, we know that $\triangle_1$ does not occur in $\mathcal{C}$.
Similarly, we can prove that $\triangle_2$ and $\triangle_3$ do not occur in $\mathcal{C}$.
\item[(II)] Assume that $\nabla=({\bf c}_1, {\bf c}_2, {\bf c}_3, {\bf c}_4, {\bf c}_5, {\bf c}_6)$ occurs in $\mathcal{C}$,
where ${\bf c}_i\in \mathcal{C}_{k_i}$ and $k_i\in\{1,2,3\}$, $i\in\{1,2,3,4,5,6\}$. Let us consider the $6$-tuple vector $(k_1,k_2,k_3,k_4,k_5,k_6)$.
\begin{itemize}
\item[(1)] Suppose that $(k_1,k_2,k_3,k_4,k_5,k_6) \in K_1 = \{(k,k,k,k,k,k)\ |\ k\in\{1,2,3\}\}$.
This implies that the codewords ${\bf c}_1, {\bf c}_2, {\bf c}_3, {\bf c}_4, {\bf c}_5, {\bf c}_6$ are all in the same subcode $\mathcal{C}_k$,
$k\in\{1,2,3\}$, a contradiction to the fact that $\mathcal{C}_k$ is a $\overline{3}$-SC.
\item [(2)] Suppose that $(k_1,k_2,k_3,k_4,k_5,k_6)\in K_2$, where $K_2$ equals\\[0.2cm]
\indent \ $\{(2,k_2,k_3,3,k_5,k_6),(2,k_2,k_3,k_4,3,k_6),(2,k_2,k_3,k_4,k_5,3),(3,k_2,k_3,2,k_5,k_6)$,\\
\indent \ $\ \ (3,k_2,k_3,k_4,2,k_6),(3,k_2,k_3,k_4,k_5,2),(k_1,2,k_3,3,k_5,k_6),(k_1,2,k_3,k_4,3,k_6)$,\\
\indent \ $\ \ (k_1,2,k_3,k_4,k_5,3),(k_1,3,k_3,2,k_5,k_6),(k_1,3,k_3,k_4,2,k_6),(k_1,3,k_3,k_4,k_5,2)$,\\
\indent \ $\ \ (k_1,k_2,2,3,k_5,k_6),(k_1,k_2,2,k_4,3,k_6),(k_1,k_2,2,k_4,k_5,3),(k_1,k_2,3,2,k_5,k_6)$,\\
\indent \ $ \ \ (k_1,k_2,3,k_4,2,k_6),(k_1,k_2,3,k_4,k_5,2)\ |\ k_1,k_2,k_3,k_4,k_5,k_6\in \{1,2,3\}\}$.\\
This contradicts the statement (iii) of Corollary \ref{co6-8}.
\item[(3)] Suppose that $(k_1,k_2,k_3,k_4,k_5,k_6) \in K_3 = \{1,2,3\}^6 \setminus (K_1\cup K_2)$.
It is easy to check that none of the above cases equals $\nabla$. We take $(1,1,1,1,2,3)$ as an example.
Let $\nabla=(\mbox{\boldmath$\alpha$}_1, \mbox{\boldmath$\alpha$}_2, \mbox{\boldmath$\alpha$}_3, \mbox{\boldmath$\alpha$}_4,
\mbox{\boldmath$\beta$}_5,\mbox{\boldmath$\gamma$}_6)$. We have\\
\indent \ \ \ \ \ \ \ \ \ \ \ \ $a_{1}r+b_{1} = a_{4}r+b_{4},$\\
\indent \ \ \ \ \ \ \ \ \ \ \ \ $a_{1}r+c_{1} = r-(x_6+1)k+z_6+(h_6+1)r,$\\
\indent \ \ \ \ \ \ \ \ \ \ \ \ $b_{1}r+c_{1} = [r-(x_{5}+1)k+y_{5}]r + [r-(x_{5}+1)k+z_{5}],$\\
\indent \ \ \ \ \ \ \ \ \ \ \ \  $a_{2}r+b_{2} = x_{5}k+y_{5}+h_{5}r,$\\
\indent \ \ \ \ \ \ \ \ \ \ \ \ $a_{2}r+c_{2} = a_{4}r+c_{4},$\\
\indent \ \ \ \ \ \ \ \ \ \ \ \ $b_{3}r+c_{3} = [r-(x_{6}+1)k+y_{6}]r + x_{6}k+z_{6},$\\
\indent \ \ \ \ \ \ \ \ \ \ \ \ $a_{3}r+b_{3} = x_{6}k+y_{6}+h_{6}r,$\\
\indent \ \ \ \ \ \ \ \ \ \ \ \ $ a_{3}r+c_{3} = x_{5}k+z_{5}+(h_{5}+1)r,$\\
\indent \ \ \ \ \ \ \ \ \ \ \ \ $ b_{3}r+c_{3} = b_{4}r+c_{4}.$\\
Then $h_6 \equiv h_6+2\ mod\ r$, a contradiction to the fact $r\geq 4$.
\end{itemize}
According to (1)-(3) above, we know that $\nabla$ does not occur in $\mathcal{C}$.
\end{itemize}

From the discussion above, we know that $\mathcal{C}$ is a $\overline{3}$-SC,
and $|\mathcal{C}|=|\mathcal{C}_1\cup\mathcal{C}_2\cup \mathcal{C}_3|=|\mathcal{C}_1|+|\mathcal{C}_2|+|\mathcal{C}_3|=r^2(k+r)$.
The proof is then completed. \qed

\subsection{Constructions via Steiner triple systems}

In this subsection, some $\overline{3}$-SC$(3,M,q)$s are constructed by means of Steiner triple systems,
which are first used to construct partial Latin squares. Let us see its definition first.

\begin{definition} \rm
\label{d21}
Let $v$ be a positive integer. A {\em Steiner triple system} (STS$(v)$ in short) is a set system $(V,\mathcal{B})$
where $V$ is a set of $v$ elements and $\mathcal{B}$ is a set of $3$-subsets of $V$ called blocks such that
every pair of distinct elements of $V$ occurs in exactly one block of $\mathcal{B}$.
\end{definition}

It is well-known that Steiner triple systems can be constructed from $(v,3,1)$-difference families.
Let $\mathcal{F}$ be a family of $3$-subsets of an additive group $G$ with order $v$.
$\mathcal{F}$ is called a {\em difference family} (briefly denoted $(v,3,1)$-DF)
if any nonzero element of $G$ can be represented in a unique way as a difference of two elements lying in some member of $\mathcal{F}$.
Then the set system $(G, \mathcal{B})$ is a Steiner triple system of order $v$, where $\mathcal{B}=\{B+g\ |\ B\in \mathcal{F}, g\in G\}$.

\begin{lemma} \rm(\cite{Wil})
\label{le1}
Let $q=6t+1$ be a prime power, $\varepsilon$ be a primitive element in $F_q$,
and $\xi={\varepsilon}^{2t}$ be a primitive $3$rd root of unity in $F_q$.
Then $\mathcal{F}=\{\{1, \xi,\xi^2\}\varepsilon^{i}\ |\  0\leq i\leq t-1\}$ is a $(v,3,1)$-DF.
\end{lemma}

Given an STS$(v)$ $(V,\mathcal{B})$, we can define its corresponding partial Latin square $\mathbb{P}$
to be the $v \times v$-array with entry $k$ in cell $(i,j)$ if and only if $\{i,j,k\}\in \mathcal{B}$, that is, we can derive
six entries $(x,y;z)$, $(x,z;y)$, $(y,x;z)$, $(y,z;x)$, $(z,y;x)$ and $(z,x;y)$ in $\mathbb{P}$ from each triple $\{x,y,z\} \in \mathcal{B}$.

Given a set $S\subseteq \{0,1,\ldots,t-1\}$, let $D=\{(1, \xi,\xi^2)^T\varepsilon^{i}\ |\  i\in S\}$ and $\mathcal{C}=\{D+g\ |\ g\in F_q\}$.
Clearly, the related array of $\mathcal{C}$ is a partial Latin square with $q|S|$ non-empty cells.
From Lemma \ref{le1319}, we know that $\mathcal{C}$ is a $2$-FPC$(3,q|S|,q)$.
Now, let us consider the forbidden configurations in (\ref{fo12}) and (\ref{fo13}).

\begin{itemize}
\item[(I)] Without loss of generality, we may assume
\begin{eqnarray*}
\label{eq20}
\begin{small}
\triangle_1=\left(\begin{array}{cccc}
\varepsilon^{x} \ \ \         & \varepsilon^{y}+k_1 \ \ \        & \varepsilon^{z}+k_2 \ \ \       & \varepsilon^{w}+k_3 \ \ \ \\
\varepsilon^{x}\xi \ \ \     & \varepsilon^{y}\xi+k_1 \ \ \     & \varepsilon^{z}\xi +k_2 \ \ \     & \varepsilon^{w}\xi+k_3 \ \ \ \\
\varepsilon^{x}\xi^{2} \ \ \  &\varepsilon^{y}\xi^{2}+k_1 \ \ \  & \varepsilon^{z}\xi^{2}+k_2 \ \ \  & \varepsilon^{w} \xi^{2}+k_3 \ \ \
\end{array}\right),\end{small}
\end{eqnarray*}
where $x,y,z,w\in S$ and $k_1$, $k_2$, $k_3\in F_q$. Then we have
\begin{eqnarray*}
\label{eq27}
\begin{small}
\left\{\begin{array}{l}
\varepsilon^{x} = \varepsilon^{y}+k_1\\
\varepsilon^{z}+k_2 = \varepsilon^{w}+k_3\\
\varepsilon^{x}\xi = \varepsilon^{w}\xi+k_3 \\
\varepsilon^{x}\xi^{2}=\varepsilon^{z}\xi^{2}+k_2\\
\varepsilon^{y}\xi^2+k_1=\varepsilon^{w}\xi^2+k_3
\end{array}
\right. \Longleftrightarrow\ \ \
\left\{\begin{array}{l}
k_1=\varepsilon^{x}-\varepsilon^{y} \\
\varepsilon^{z}+\varepsilon^{x}\xi^{2}-\varepsilon^{z}\xi^{2} = \varepsilon^{w}+\varepsilon^{x}\xi-\varepsilon^{w}\xi \\
k_3=\varepsilon^{x}\xi-\varepsilon^{w}\xi \\
k_2= \varepsilon^{x}\xi^{2}-\varepsilon^{z}\xi^{2} \\
\varepsilon^{y}\xi^2+\varepsilon^{x}-\varepsilon^{y}= \varepsilon^{w}\xi^2+ \varepsilon^{x}\xi-\varepsilon^{w}\xi
\end{array}
\right.
\end{small}
\end{eqnarray*}
This means \begin{eqnarray*}
\label{eq39}
\varepsilon^{w}+\varepsilon^{x}\xi+\varepsilon^{z}\xi^2=0 \ \ \hbox{and}\ \
\varepsilon^{x}+\varepsilon^{w}\xi+\varepsilon^{y}\xi^2=0,
\end{eqnarray*}
with $|\{x, y, z,w\}|=4$. In fact,
\begin{enumerate}
\item[(1)] $x\neq y,z,w$ always holds. If $x=y$, we have $k_1=\varepsilon^{x}-\varepsilon^{y}=0$.
This implies that the first codeword equals the second codeword of $\triangle_1$, a contradiction to the assumption.
Similarly, from $k_2= \varepsilon^{x}\xi^{2}-\varepsilon^{z}\xi^{2}$ and $k_3=\varepsilon^{x}\xi-\varepsilon^{w}\xi$, we have $x\neq z,w$.
\item [(2)] $y\neq z,w$ always holds. If $y=z$, we have $x=w$ from $\varepsilon^{w}+\varepsilon^{x}\xi+\varepsilon^{z}\xi^2=0$ and
$\varepsilon^{x}+\varepsilon^{w}\xi+\varepsilon^{y}\xi^2=0$. This is a contradiction to the fact $x\neq w$.
Similarly, from $\varepsilon^{x}+\varepsilon^{w}\xi+\varepsilon^{y}\xi^2=0$, we have $y\neq w$.
\item [(3)] $z\neq w$ always holds. If $z=w$, from $\varepsilon^{w}+\varepsilon^{x}\xi+\varepsilon^{z}\xi^2=0$, we have $x=z$,
a contradiction to the fact $x\neq z$.
\end{enumerate}
The converse can be also proved to be true.

From the discussion above, we know that $\triangle_1$ occurs in $\mathcal{C}$
if and only if there exists a solution to the following system of equations.
\begin{eqnarray}
\label{eq25}
\begin{small}
\left\{\begin{array}{l}
\varepsilon^{w}+\varepsilon^{x}\xi+\varepsilon^{z}\xi^2=0\\[0.1cm]
\varepsilon^{x}+\varepsilon^{w}\xi+\varepsilon^{y}\xi^2=0\\[0.1cm]
|\{x,y,z,w\}|=4
\end{array},
\right.\end{small} \ \hbox{where}\ x,y,z,w\in S.
\end{eqnarray}

Similarly, we can check that any of $\triangle_2$, $\triangle_3$ occurs in $\mathcal{C}$
if and only if there exists a solution to the above system of equations.
\item[(II)] Without loss of generality, we may assume
\begin{eqnarray*}
\label{eq21}\begin{small}
\nabla=\left(\begin{array}{cccccc}
\varepsilon^{x}\ \ \         & \varepsilon^{y}+k_1\ \ \        & \varepsilon^{z}+k_2\ \ \          & \varepsilon^{u}+k_3\ \ \       & \varepsilon^{v}+k_4\ \ \       & \varepsilon^{w}+k_5\ \ \ \\
\varepsilon^{x}\xi \ \ \     & \varepsilon^{y}\xi+k_1\ \ \     & \varepsilon^{z}\xi +k_2\ \ \      & \varepsilon^{u}\xi+k_3\ \ \    & \varepsilon^{v}\xi+k_4\ \ \    & \varepsilon^{w}\xi+k_5\ \ \ \\
\varepsilon^{x}\xi^{2}\ \ \  &\varepsilon^{y}\xi^{2}+k_1\ \ \  & \varepsilon^{z}\xi^{2}+k_2\ \ \   & \varepsilon^{u}\xi^{2}+k_3\ \ \    & \varepsilon^{v}\xi^{2}+k_4\ \ \    & \varepsilon^{w}\xi^{2}+k_5\ \ \
\end{array}\right),\end{small}
\end{eqnarray*}
where $x, y, z,u,v,w\in S$ and $k_1,k_2,k_3, k_4,k_5\in F_q$. Then we have
\begin{eqnarray*}
\label{eq23}
\begin{small}
\left\{\begin{array}{lllll}
\varepsilon^{x}=\varepsilon^{u}+k_3\\
\varepsilon^{y}+k_1=\varepsilon^{v}+k_4\\
\varepsilon^{z}+k_2=\varepsilon^{w}+k_5\\
\varepsilon^{x}\xi=\varepsilon^{w}\xi+k_5\\
\varepsilon^{y}\xi+k_1=\varepsilon^{u}\xi+k_3\\
\varepsilon^{z}\xi +k_2=\varepsilon^{v}\xi+k_4\\
\varepsilon^{x}\xi^2=\varepsilon^{v}\xi^{2}+k_4\\
\varepsilon^{y}\xi^2+k_1=\varepsilon^{w}\xi^{2}+k_5\\
\varepsilon^{z}\xi^2+k_2=\varepsilon^{u}\xi^{2}+k_3
\end{array}
\right.\Longleftrightarrow
\left\{\begin{array}{lllll}
k_3=\varepsilon^{x}-\varepsilon^{u}\\
k_1=\varepsilon^{v}+\varepsilon^{x}\xi^2-\varepsilon^{v}\xi^{2}-\varepsilon^{y}\\
k_2=\varepsilon^{w}+\varepsilon^{x}\xi-\varepsilon^{w}\xi-\varepsilon^{z}\\
k_5=\varepsilon^{x}\xi-\varepsilon^{w}\xi \\
k_1=\varepsilon^{u}\xi+\varepsilon^{x}-\varepsilon^{u}-\varepsilon^{y}\xi \\
k_2=\varepsilon^{v}\xi+\varepsilon^{x}\xi^2-\varepsilon^{v}\xi^{2}-\varepsilon^{z}\xi \\
k_4=\varepsilon^{x}\xi^2-\varepsilon^{v}\xi^{2}\\
k_1=\varepsilon^{w}\xi^{2}+\varepsilon^{x}\xi-\varepsilon^{w}\xi-\varepsilon^{y}\xi^2\\
k_2=\varepsilon^{u}\xi^{2}+\varepsilon^{x}-\varepsilon^{u}-\varepsilon^{z}\xi^2
\end{array}.
\right.
\end{small}
\end{eqnarray*}
This means
\begin{eqnarray*}
\label{eq32}
\varepsilon^{x}+ \varepsilon^{y}\xi^2+\varepsilon^{z}\xi=0,\
\varepsilon^{u}+\varepsilon^{v}\xi+\varepsilon^{w}\xi^2=0\ \hbox{and}\
\varepsilon^{x}\xi+\varepsilon^{u}
   =\varepsilon^{z}+\varepsilon^{w}\xi,
\end{eqnarray*}
\noindent where
\begin{eqnarray}
\left\{\begin{array}{lllll}
x \not\in \{u,v,w\} \ \hbox{and} \ |\{u,v,w\}|=3 \ & \hbox{if} \ x=y=z;\\
u \not\in \{x,y,z\} \ \hbox{and} \ |\{x,y,z\}|=3 \ & \hbox{if} \ u=v=w;\\
|\{x,y,z,u,v,w\}|=6 & \hbox{otherwise.}
\end{array}
\right.
\end{eqnarray}
From the discussion above, we have
\begin{enumerate}
\item[(1)] $\{x,y,z\}\cap \{u,v,w\}=\emptyset$ always holds.
\subitem$\bullet$ $x\not\in\{u,v,w\}$ always holds. If $x=u$, we have $k_3=\varepsilon^{x}-\varepsilon^{u}=0$.
This implies that the first codeword of $\nabla$ equals the forth codeword of $\nabla$, a contradiction to the assumption.
Similarly, from $k_5=\varepsilon^{x}\xi-\varepsilon^{w}\xi$ and $k_4=\varepsilon^{x}\xi^2-\varepsilon^{v}\xi^{2}$,
we can prove that $x \neq w,v$ always holds.
\subitem$\bullet$ $y\not\in\{u,v,w\}$ always holds. If $y=u$, we have $k_1=k_3$ from $\varepsilon^{y}\xi+k_1=\varepsilon^{u}\xi+k_3$.
This implies that the second codeword of $\nabla$ equals the forth codeword of $\nabla$, a contradiction to the assumption.
Similarly, from $\varepsilon^{y}+k_1=\varepsilon^{v}+k_4$ and $\varepsilon^{y}\xi^2+k_1=\varepsilon^{w}\xi^{2}+k_5$, we have $y\neq v,w$.
\subitem$\bullet$ $z\not\in\{u,v,w\}$ always holds. If $z=u$, we have $k_2=k_3$ from $\varepsilon^{z}\xi^2+k_2=\varepsilon^{u}\xi^{2}+k_3$.
This implies that the third codeword of $\nabla$ equals the forth codeword of $\nabla$, a contradiction to the assumption.
Similarly, from $\varepsilon^{z}+k_2=\varepsilon^{w}+k_5$ and $\varepsilon^{z}\xi +k_2=\varepsilon^{v}\xi+k_4$, we have $z \neq w,v$.
\item [(2)] If $|\{x,y,z\}|<3$ (or $|\{u,v,w\}|<3$), then $|\{x,y,z\}|=1$ (or $|\{u,v,w\}|=1$) always holds.
Without loss of generality, we may assume $x=y$. Then we have $x=y=z$ from $\varepsilon^{x}+ \varepsilon^{y}\xi^2+\varepsilon^{z}\xi=0$.
Similarly, we have that if $|\{u,v,w\}|<3$, then $|\{u,v,w\}|=1$ always holds from $\varepsilon^{u}+\varepsilon^{v}\xi+\varepsilon^{w}\xi^2=0$.
\item [(3)] If $x=y=z$ (or $u=v=w$), then $|\{u,v,w\}|=3$ (or $|\{x,y,z\}|=3$) always holds.
If $x=y=z$ and $u=v=w$, we have $\varepsilon^{x} =\varepsilon^{u}$ from $\varepsilon^{x}\xi+\varepsilon^{u} =\varepsilon^{z}+\varepsilon^{w}\xi$.
This is a contradiction to the fact $x\neq u$.
\end{enumerate}
The converse can be also proved to be true.

Summarizing the discussion above, we know that $\nabla$ occurs in $\mathcal{C}$ if and only if
there exists a solution to the following system of equations.
\begin{eqnarray}
\label{eq41}
\left\{\begin{array}{lllll}
\varepsilon^{x}+ \varepsilon^{y}\xi^2+\varepsilon^{z}\xi=0\\
\varepsilon^{u}+\varepsilon^{v}\xi+\varepsilon^{w}\xi^2=0\\
\varepsilon^{x}\xi+\varepsilon^{u}
   =\varepsilon^{z}+\varepsilon^{w}\xi
\end{array},
\right. \hbox{where}
\left\{\begin{array}{lllll}
x\not\in \{u,v,w\} \ \hbox{and}\ |\{u,v,w\}|=3\ & \hbox{if}\ x=y=z;\\
u\not\in \{x,y,z\} \ \hbox{and}\ |\{x,y,z\}|=3\ & \hbox{if}\ u=v=w;\\
|\{x,y,z,u,v,w\}|=6 & \hbox{otherwise.}
\end{array}
\right.
\end{eqnarray}
\end{itemize}

Therefore we have the following result.
\begin{theorem}\rm
\label{thl}
$\mathcal{C}$ is a $\overline{3}$-SC$(3,q|S|,q)$ if and only if each solution to (\ref{eq25}) and (\ref{eq41}) does not belong to $S$.
\end{theorem}

With the aid of a computer, by applying Theorem \ref{thl}, we obtain the following results for small $q$.
\begin{corollary}\rm
\label{cor1}
\begin{enumerate}
\item[(1)] There exists a $\overline{3}$-SC$(3,\frac{q(q-1)}{6},q)$, when $q=73$, $79$, $103$, $127$, $139$.
\item[(2)] There exists a $\overline{3}$-SC$(3,\lfloor\frac{q(q-1)}{9}\rfloor,q)$, when $q=109$, $121$, $157$, $169$, $199$, $229$, $313$.
\item[(3)] There exists a $\overline{3}$-SC$(3,\lfloor\frac{q(q-1)}{12}\rfloor,q)$, when $q=151$, $157$, $163$, $193$, $211$, $223$, $241$,
$271$, $277$, $283$, $307$, $337$, $349$, $367$, $409$, $433$, $499$, $523$, $547$, $571$, $577$, $601$, $727$, $739$, $811$, $859$.
\item[(4)] There exists a $\overline{3}$-SC$(3,\lfloor\frac{q(q-1)}{18}\rfloor,q)$, when $q=181$, $313$, $331$, $343$, $373$,
$397$, $421$, $439$, $457$, $499$, $529$, $541$, $607$, $613$, $619$, $625$, $673$, $691$, $733$, $751$, $757$, $787$, $823$, $841$, $877$, $907$, $919$, $937$, $961$,
$967$, $991$, $997$.
\end{enumerate}
\end{corollary}
{\bf Proof:} The results are obtained by letting $S = \{0,1,\ldots,t-1\}$, $\{i\ |\ i\not\equiv 0 \ mod\ 3, 0\leq i<t\}$,
$\{i\ |\ i\equiv 0\ mod\ 2, 0\leq i<t\}$, $\{i\ |\ i\equiv 0\ mod\ 3, 0 \le i < t\}$, respectively.
\qed

\section{Summary}   %

In this paper, a new upper bound $M(\overline{3},3,q)\leq \lfloor\frac{3q^2}{4}\rfloor$ was derived
by considering an optimization problem related to a partial Latin square.
Taking advantage of perfect hash families, we constructed $\overline{3}$-SC$(3,r^3,r^2)$s for all positive integers $r$.
Consequently, a lower bound $M(\overline{3},3,q)\geq\lfloor\sqrt{q}\rfloor^3$ was obtained,
and for every even integer $k$, a $\overline{3}$-SC$(3,r^3+kr^2,r^2)$ was constructed where $r=k^2$.
Finally, the constructions of $\overline{3}$-SC$(3,M,q)$ from Steiner triple systems were also discussed.


\section*{Appendix: Proof of Theorem~\ref{th2610}}
The necessity is clear from Theorem~\ref{adle1} and the discussion in Section~\ref{forb}. We consider its sufficiency.
From the relationship between a separable code and a frameproof code in Theorem~\ref{adle1}, it is sufficient to prove that
for any distinct $A, B\subseteq \mathcal{C}$ with $|A|=|B|=3$, we have ${\sf desc}(A)\neq {\sf desc}(B)$
except that $A \bigcup B$ equals one of the forbidden configurations in (\ref{fo12}) and (\ref{fo13}).

Let $A=\{{\bf a}=(a_1, a_2, a_3)^T$, ${\bf b}=(b_1,b_2,b_3)^T$, ${\bf c}=(c_1,c_2,c_3)^T\}\subseteq \mathcal{C}$ and $B=\{{\bf d}=(d_1,d_2,d_3)^T$, ${\bf e}=(e_1,e_2,e_3)^T$,
${\bf f}=(f_1,f_2,f_3)^T\}$ $\subseteq \mathcal{C}$. Suppose that ${\sf desc}(A)={\sf desc}(B)$,
$A\neq B$ and $|A|=|B|=3$. Then we have $\{a_1,b_1,c_1\}=\{d_1,e_1,f_1\}$, $\{a_2,b_2,c_2\}=\{d_2,e_2,f_2\}$
and $\{a_3,b_3,c_3\}=\{d_3,e_3,f_3\}$ by the definition of a separable code. There are three cases to be considered:\\[0.25cm]
(I) $|A\bigcap B|=2$. Without loss of generality, we may assume that ${\bf b}={\bf e}$ and ${\bf c}={\bf f}$,
that is, $A=\{{\bf a},{\bf b},{\bf c}\}$, $B=\{{\bf b},{\bf c},{\bf d}\}$. We list them as follows.
\begin{eqnarray}
\label{eq04}
\left(\begin{array}{cccc}
a_{1} & b_{1} & c_{1} & d_{1}\\
a_{2} & b_{2} & c_{2} & d_{2}\\
a_{3} & b_{3} & c_{3} & d_{3}
\end{array}\right)
\end{eqnarray}Since $\mathcal{C}$ is a $2$-FPC, the following statements hold.
\subitem\ding{227} ${\bf d}\not\in {\sf desc}(\{{\bf a},{\bf b}\})$ holds because ${\bf d}\not\in \{{\bf a},{\bf b}\}$.
Without loss of generality, we may assume that $d_{1}\not\in \{a_{1},b_{1}\}$,
then $c_1=d_1$ and $a_1=b_1$ always hold since $\{a_1,b_1,c_1\}=\{b_1,c_1,d_1\}$.
Then (\ref{eq04}) can be written as follows.
\begin{eqnarray}
\label{eq05}
\left(\begin{array}{cccc}
a_{1} & a_{1} & c_{1} & c_{1}\\
a_{2} & b_{2} & c_{2} & d_{2}\\
a_{3} & b_{3} & c_{3} & d_{3}
\end{array}\right)
\end{eqnarray}
\subitem\ding{227} ${\bf d}\not\in {\sf desc}(\{{\bf b},{\bf c}\})$ holds because ${\bf d}\not\in \{{\bf b},{\bf c}\}$.
Without loss of generality, we may assume that $d_{2}\not\in \{b_{2},c_{2}\}$, then $a_2=d_2$ always holds
since $\{a_2,b_2,c_2\}=\{b_2,c_2,d_2\}$.
Then (\ref{eq05}) can be written as follows.
\begin{eqnarray}
\label{eq06}
\left(\begin{array}{cccc}
a_{1} & a_{1} & c_{1} & c_{1}\\
a_{2} & b_{2} & c_{2} & a_{2}\\
a_{3} & b_{3} & c_{3} & d_{3}
\end{array}\right)
\end{eqnarray}
\subitem\ding{227} ${\bf d}\not\in {\sf desc}(\{{\bf a},{\bf c}\})$ holds because ${\bf d}\not\in \{{\bf a},{\bf c}\}$.
Then we must have $d_{3}\not\in \{a_{3},c_{3}\}$, which implies $b_3=d_3$ and $a_3=c_3$ since $\{a_3,b_3,c_3\}=\{b_3,c_3,d_3\}$.
Then (\ref{eq06}) can be written as follows.
\begin{eqnarray}
\label{eq07}
\left(\begin{array}{cccc}
a_{1} & a_{1} & c_{1} & c_{1}\\
a_{2} & b_{2} & c_{2} & a_{2}\\
a_{3} & b_{3} & a_{3} & b_{3}
\end{array}\right)
\end{eqnarray}
Obviously, (\ref{eq07}) equals $\triangle_1$ in (\ref{fo12}), which is a forbidden configuration in $\mathcal{C}$.

\noindent(II) $|A\bigcap B|=1$. Without loss of generality, we may assume that ${\bf c}={\bf f}$, that is,
$A=\{{\bf a},{\bf b},{\bf c}\}$, $B=\{{\bf c},{\bf d},{\bf e}\}$. We list them as follows.
\begin{eqnarray}
\label{eq09}
\left(\begin{array}{ccccc}
a_{1} & b_{1} & c_{1} & d_{1} & e_{1}\\
a_{2} & b_{2} & c_{2} & d_{2} & e_{2}\\
a_{3} & b_{3} & c_{3} & d_{3} & e_{3}
\end{array}\right)
\end{eqnarray}Since $\mathcal{C}$ is a $2$-FPC, the following statements hold.
\subitem\ding{227} ${\bf d}\not\in {\sf desc}(\{{\bf a},{\bf b}\})$ holds because ${\bf d}\not\in \{{\bf a},{\bf b}\}$.
Without loss of generality, we may assume that $d_{1}\not\in \{a_{1},b_{1}\}$, then $c_1=d_1$ and $a_1=b_1=e_1$ always hold
since $\{a_1,b_1,c_1\}=\{c_1,d_1,e_1\}$.
Then (\ref{eq09}) can be written as follows.
\begin{eqnarray}
\label{eq10}
\left(\begin{array}{ccccc}
a_{1} & a_{1} & c_{1} & c_{1} & a_{1}\\
a_{2} & b_{2} & c_{2} & d_{2} & e_{2}\\
a_{3} & b_{3} & c_{3} & d_{3} & e_{3}
\end{array}\right)
\end{eqnarray}
\ding{227}  ${\bf e}\not\in {\sf desc}(\{{\bf a},{\bf b}\})$ holds because ${\bf e}\not\in \{{\bf a},{\bf b}\}$.
Without loss of generality, we may assume that $e_{2}\not\in \{a_{2},b_{2}\}$, then $c_2=e_2$ and $a_2=b_2=d_2$ always hold
since $\{a_2,b_2,c_2\}=\{c_2,d_2,e_2\}$.
Then (\ref{eq10}) can be written as follows.
\begin{eqnarray}
\label{eq11}
\left(\begin{array}{ccccc}
a_{1} & a_{1} & c_{1} & c_{1} & a_{1}\\
a_{2} & a_{2} & c_{2} & a_{2} & c_{2}\\
a_{3} & b_{3} & c_{3} & d_{3} & e_{3}
\end{array}\right)
\end{eqnarray}
\ding{227} ${\bf d}\not\in {\sf desc}(\{{\bf b},{\bf c}\})$ holds because ${\bf d}\not\in \{{\bf b},{\bf c}\}$.
Then we must have $d_{3}\not\in \{b_{3},c_{3}\}$, which implies $a_3=d_3$ since $\{a_3,b_3,c_3\}=\{c_3,d_3,e_3\}$.
Then (\ref{eq11}) can be written as follows.
\begin{eqnarray*}
\label{eq12}
\left(\begin{array}{ccccc}
a_{1} & a_{1} & c_{1} & c_{1} & a_{1}\\
a_{2} & a_{2} & c_{2} & a_{2} & c_{2}\\
a_{3} & b_{3} & c_{3} & a_{3} & e_{3}
\end{array}\right)
\end{eqnarray*}
Obviousely,  ${\bf d}\in {\sf desc}(\{{\bf a},{\bf c}\})$ holds.
This contradicts the definition of a $2$-FPC since ${\bf d}\not\in \{{\bf a},{\bf c}\}$.

So the case of $|A\bigcap B|=1$ can never happen.

\noindent(III) $|A\bigcap B|=0$.  We list $A$ and $B$ as follows.
\begin{eqnarray}
\label{eq14}
\left(\begin{array}{cccccc}
a_{1} & b_{1} & c_{1} & d_{1} & e_{1} & f_{1}\\
a_{2} & b_{2} & c_{2} & d_{2} & e_{2} & f_{2}\\
a_{3} & b_{3} & c_{3} & d_{3} & e_{3} & f_{3}
\end{array}\right)
\end{eqnarray}Since $\mathcal{C}$ is a $2$-FPC, the following statements hold.
\subitem\ding{227} ${\bf d}\not\in {\sf desc}(\{{\bf a},{\bf b}\})$ holds because ${\bf d}\not\in \{{\bf a},{\bf b}\}$. Without loss of generality, we may assume that $d_{1}\not\in \{a_{1},b_{1}\}$, then $c_1=d_1$ always holds since $\{a_1,b_1,c_1\}=\{d_1,e_1,f_1\}$.
Then (\ref{eq14}) can be written as follows.
\begin{eqnarray}
\label{eq15}
\left(\begin{array}{cccccc}
a_{1} & b_{1} & c_{1} & c_{1} & e_{1} & f_{1}\\
a_{2} & b_{2} & c_{2} & d_{2} & e_{2} & f_{2}\\
a_{3} & b_{3} & c_{3} & d_{3} & e_{3} & f_{3}
\end{array}\right)
\end{eqnarray}
\subitem\ding{227} ${\bf d}\not\in {\sf desc}(\{{\bf a},{\bf c}\})$ holds because ${\bf d}\not\in \{{\bf a},{\bf c}\}$. Without loss of generality, we may assume that $d_{2}\not\in \{a_{2},c_{2}\}$, then $b_2=d_2$ always holds since $\{a_2,b_2,c_2\}=\{d_2,e_2,f_2\}$.
Then (\ref{eq15}) can be written as follows.
\begin{eqnarray}
\label{eq16}
\left(\begin{array}{cccccc}
a_{1} & b_{1} & c_{1} & c_{1} & e_{1} & f_{1}\\
a_{2} & b_{2} & c_{2} & b_{2} & e_{2} & f_{2}\\
a_{3} & b_{3} & c_{3} & d_{3} & e_{3} & f_{3}
\end{array}\right)
\end{eqnarray}
\subitem\ding{227} ${\bf d}\not\in {\sf desc}(\{{\bf b},{\bf c}\})$ holds because ${\bf d}\not\in \{{\bf b},{\bf c}\}$. Then we must have $d_{3}\not\in \{b_{3},c_{3}\}$, which implies $a_3=d_3$ since $\{a_3,b_3,c_3\}=\{d_3,e_3,f_3\}$.
Then (\ref{eq16}) can be written as follows.
\begin{eqnarray*}
\label{eq17}
\left(\begin{array}{cccccc}
a_{1} & b_{1} & c_{1} & c_{1} & e_{1} & f_{1}\\
a_{2} & b_{2} & c_{2} & b_{2} & e_{2} & f_{2}\\
a_{3} & b_{3} & c_{3} & a_{3} & e_{3} & f_{3}
\end{array}\right)
\end{eqnarray*}
\subitem\ding{227} ${\bf e}\not\in {\sf desc}(\{{\bf a},{\bf b}\})$ holds because ${\bf e}\not\in \{{\bf a},{\bf b}\}$. There must exist an index $1\leq i\leq 3$ such that $e_{i}\not\in \{a_{i},b_{i}\}$ holds.\\
\indent \ \ (1.1) If $e_{1}\not\in \{a_{1},b_{1}\}$ holds, then $c_1=e_1$ always holds since $\{a_1,b_1,c_1\}=\{c_1,e_1,f_1\}$;\\
\indent \ \ (1.2) If $e_{2}\not\in \{a_{2},b_{2}\}$ holds, then $c_2=e_2$ always holds since $\{a_2,b_2,c_2\}=\{b_2,e_2,f_2\}$;\\
\indent \ \ (1.3) If $e_{3}\not\in \{a_{3},b_{3}\}$ holds, then $c_3=e_3$ always holds since $\{a_3,b_3,c_3\}=\{a_3,e_3,f_3\}$.
\subitem\ding{227} ${\bf e}\not\in {\sf desc}(\{{\bf a},{\bf c}\})$ holds because ${\bf e}\not\in \{{\bf a},{\bf c}\}$. There must exist an integer $1\leq i\leq 3$ such that $e_{i}\not\in \{a_{i},c_{i}\}$ holds.\\
\indent \ \ (2.1) If $e_{1}\not\in \{a_{1},c_{1}\}$ holds, then $b_1=e_1$ always holds since $\{a_1,b_1,c_1\}=\{c_1,e_1,f_1\}$;\\
\indent \ \ (2.2) If $e_{2}\not\in \{a_{2},c_{2}\}$ holds, then $b_2=e_2$ always holds since $\{a_2,b_2,c_2\}=\{b_2,e_2,f_2\}$;\\
\indent \ \ (2.3) If $e_{3}\not\in \{a_{3},c_{3}\}$ holds, then $b_3=e_3$ always holds since $\{a_3,b_3,c_3\}=\{a_3,e_3,f_3\}$.
\subitem\ding{227} ${\bf e}\not\in {\sf desc}(\{{\bf b},{\bf c}\})$ holds because ${\bf e}\not\in \{{\bf b},{\bf c}\}$. There must exist an index $1\leq i\leq 3$ such that $e_{i}\not\in \{b_{i},c_{i}\}$ holds.\\
\indent \ \ (3.1) If $e_{1}\not\in \{b_{1},c_{1}\}$ holds, then $a_1=e_1$ always holds since $\{a_1,b_1,c_1\}=\{c_1,e_1,f_1\}$;\\
\indent \ \ (3.2) If $e_{2}\not\in \{b_{2},c_{2}\}$ holds, then $a_2=e_2$ always holds since $\{a_2,b_2,c_2\}=\{b_2,e_2,f_2\}$;\\
\indent \ \ (3.3) If $e_{3}\not\in \{b_{3},c_{3}\}$ holds, then $a_3=e_3$ always holds since $\{a_3,b_3,c_3\}=\{a_3,e_3,f_3\}$.
\subitem\ding{227} ${\bf f}\not\in {\sf desc}(\{{\bf a},{\bf b}\})$ holds because ${\bf f}\not\in \{{\bf a},{\bf b}\}$. There must exist an index $1\leq i\leq 3$ such that $f_{i}\not\in \{a_{i},b_{i}\}$ holds.\\
\indent \ \ (4.1) If $f_{1}\not\in \{a_{1},b_{1}\}$ holds, then $c_1=f_1$ always holds since $\{a_1,b_1,c_1\}=\{c_1,e_1,f_1\}$;\\
\indent \ \ (4.2) If $f_{2}\not\in \{a_{2},b_{2}\}$ holds, then $c_2=f_2$ always holds since $\{a_2,b_2,c_2\}=\{b_2,e_2,f_2\}$;\\
\indent \ \ (4.3) If $f_{3}\not\in \{a_{3},b_{3}\}$ holds, then $c_3=f_3$ always holds since $\{a_3,b_3,c_3\}=\{a_3,e_3,f_3\}$.
\subitem\ding{227} ${\bf f}\not\in {\sf desc}(\{{\bf a},{\bf c}\})$ holds because ${\bf f}\not\in \{{\bf a},{\bf c}\}$. There must exist an index $1\leq i\leq 3$ such that $f_{i}\not\in \{a_{i},c_{i}\}$ holds.\\
\indent \ \ (5.1) If $f_{1}\not\in \{a_{1},c_{1}\}$ holds, then $b_1=f_1$ always holds since $\{a_1,b_1,c_1\}=\{c_1,e_1,f_1\}$;\\
\indent \ \ (5.2) If $f_{2}\not\in \{a_{2},c_{2}\}$ holds, then $b_2=f_2$ always holds since $\{a_2,b_2,c_2\}=\{b_2,e_2,f_2\}$;\\
\indent \ \ (5.3) If $f_{3}\not\in \{a_{3},c_{3}\}$ holds, then $b_3=f_3$ always holds since $\{a_3,b_3,c_3\}=\{a_3,e_3,f_3\}$.
\subitem\ding{227} ${\bf f}\not\in {\sf desc}(\{{\bf b},{\bf c}\})$ holds because ${\bf f}\not\in \{{\bf b},{\bf c}\}$. There must exist an index $1\leq i\leq 3$ such that $f_{i}\not\in \{b_{i},c_{i}\}$ holds.\\
\indent \ \ (6.1) If $f_{1}\not\in \{b_{1},c_{1}\}$ holds, then $a_1=f_1$ always holds since $\{a_1,b_1,c_1\}=\{c_1,e_1,f_1\}$;\\
\indent \ \ (6.2) If $f_{2}\not\in \{b_{2},c_{2}\}$ holds, then $a_2=f_2$ always holds since $\{a_2,b_2,c_2\}=\{b_2,e_2,f_2\}$;\\
\indent \ \ (6.3) If $f_{3}\not\in \{b_{3},c_{3}\}$ holds, then $a_3=f_3$ always holds since $\{a_3,b_3,c_3\}=\{a_3,e_3,f_3\}$.

It is easy to check that for any $i\in \{1,2,3\}$ or any $i\in \{4,5,6\}$, and for any $j\in \{1,2,3\}$, once $(i.j)$ occurs, no $(i'.j)$ can occur for any $i'\in\{1,2,3\}\setminus \{i\}$ or any $i'\in\{4,5,6\}\setminus \{i\}$, respectively. So there are $3\times 2\times 1\times 3\times 2\times 1=36$ cases to be considered.
\begin{center}
$\{(1.3),(2.2),(3.1),(4.2),(5.1),(6.3)\}$,\ \ \
$\{(1.1),(2.3),(3.2),(4.2),(5.1),(6.3)\}$,\\
$\{(1.1),(2.3),(3.2),(4.2),(5.3),(6.1)\}$,\ \ \
$\{(1.1),(2.3),(3.2),(4.3),(5.1),(6.2)\}$,\\
$\{(1.1),(2.3),(3.2),(4.3),(5.2),(6.1)\}$,\ \ \
$\{(1.2),(2.1),(3.3),(4.1),(5.3),(6.2)\}$,\\
$\{(1.2),(2.1),(3.3),(4.2),(5.3),(6.1)\}$,\ \ \
$\{(1.2),(2.1),(3.3),(4.3),(5.1),(6.2)\}$,\\
$\{(1.2),(2.1),(3.3),(4.3),(5.2),(6.1)\}$,\ \ \
$\{(1.2),(2.3),(3.1),(4.1),(5.3),(6.2)\}$,\\
$\{(1.2),(2.3),(3.1),(4.2),(5.1),(6.3)\}$,\ \ \
$\{(1.2),(2.3),(3.1),(4.3),(5.2),(6.1)\}$,\\
$\{(1.3),(2.1),(3.2),(4.1),(5.3),(6.2)\}$,\ \ \
$\{(1.3),(2.1),(3.2),(4.2),(5.1),(6.3)\}$,\\
$\{(1.3),(2.1),(3.2),(4.3),(5.2),(6.1)\}$,\ \ \
$\{(1.3),(2.2),(3.1),(4.1),(5.3),(6.2)\}$,\\
$\{(1.3),(2.2),(3.1),(4.2),(5.3),(6.1)\}$,\ \ \
$\{(1.3),(2.2),(3.1),(4.3),(5.1),(6.2)\}$,\\
$\{(1.1),(2.2),(3.3),(4.1),(5.2),(6.3)\}$,\ \ \
$\{(1.1),(2.2),(3.3),(4.1),(5.3),(6.2)\}$,\\
$\{(1.1),(2.2),(3.3),(4.2),(5.1),(6.3)\}$,\ \ \
$\{(1.1),(2.2),(3.3),(4.2),(5.3),(6.1)\}$,\\
$\{(1.1),(2.2),(3.3),(4.3),(5.1),(6.2)\}$,\ \ \
$\{(1.1),(2.2),(3.3),(4.3),(5.2),(6.1)\}$,\\
$\{(1.1),(2.3),(3.2),(4.1),(5.2),(6.3)\}$,\ \ \
$\{(1.1),(2.3),(3.2),(4.1),(5.3),(6.2)\}$,\\
$\{(1.2),(2.1),(3.3),(4.1),(5.2),(6.3)\}$,\ \ \
$\{(1.2),(2.1),(3.3),(4.2),(5.1),(6.3)\}$,\\
$\{(1.2),(2.3),(3.1),(4.1),(5.2),(6.3)\}$,\ \ \
$\{(1.2),(2.3),(3.1),(4.2),(5.3),(6.1)\}$,\\
$\{(1.3),(2.1),(3.2),(4.1),(5.2),(6.3)\}$,\ \ \
$\{(1.3),(2.1),(3.2),(4.3),(5.1),(6.2)\}$,\\
$\{(1.3),(2.2),(3.1),(4.1),(5.2),(6.3)\}$,\ \ \
$\{(1.3),(2.2),(3.1),(4.3),(5.2),(6.1)\}$,\\
$\{(1.2),(2.3),(3.1),(4.3),(5.1),(6.2)\}$,\ \ \
$\{(1.3),(2.1),(3.2),(4.2),(5.3),(6.1)\}$.
\end{center}
It is readily checked that none of the first $18$ subcases satisfies the condition (i) in this theorem.
For example, consider the subcase $\{(1.3),(2.2),(3.1),(4.2),(5.1),(6.3)\}$, that is,
$$
(1.3) \ c_3=e_3; \ (2.2) \ b_2 = e_2; \ (3.1) \ a_1 = e_1;\
(4.2) \ c_2=f_2; \ (5.1) \ b_1 = f_1; \ (6.3) \ a_3 = f_3.
$$
Then the corresponding subcode can be written as follows.
\begin{eqnarray*}
\label{eq37}
\left(\begin{array}{cccccc}
a_{1} & b_{1} & c_{1} & c_{1} & a_{1} & b_{1}\\
a_{2} & b_{2} & c_{2} & b_{2} & b_{2} & c_{2}\\
a_{3} & b_{3} & c_{3} & a_{3} & c_{3} & a_{3}
\end{array}\right)
\end{eqnarray*}
where ${\bf e}=(a_{1},b_{2},c_{3})^T$ and ${\bf f}=(b_{1},c_{2},a_{3})^T$. Obviously,
${\bf b}\in {\sf desc}(\{{\bf e},{\bf f}\})$ holds since $b_{3}\in \{a_3,c_3\}$. This is a contradiction to the definition of a $2$-FPC because of the assumption ${\bf b}\not\in \{{\bf e},{\bf f}\}$.
It is also easy to check that none of the next $16$ subcases satisfies the condition that $|B|=3$.
Finally the remaining $2$ subcases correspond to the forbidden configuration in (\ref{fo13}).

The proof is then completed.\qed

\end{document}